%% file: hetrl_mlsys2026.tex
\documentclass{article}

\usepackage{microtype}
\usepackage{graphicx}
\usepackage{subfigure}
\usepackage{booktabs} % for professional tables

\usepackage{hyperref}

\usepackage[accepted]{mlsys2025}
\usepackage{amsmath}
\usepackage{amssymb}
\usepackage{amsthm}
\theoremstyle{definition}
\newtheorem{definition}{Definition}

\usepackage{enumitem}
\usepackage{pifont}
\newcommand{\circled}[2][1.2]{%
  \raisebox{-0.3ex}{\scalebox{#1}{\ding{\numexpr 181 + #2\relax}}}%
}
\theoremstyle{plain}
\newtheorem{proposition}{Proposition}
\usepackage{xurl}
\usepackage{xspace}
\newcommand{\sys}{HetRL\xspace}

\mlsystitlerunning{\sys: Efficient RL for LLMs in Heterogeneous Environments}

\begin{document}

\twocolumn[
\mlsystitle{\sys: Efficient Reinforcement Learning for LLMs in Heterogeneous Environments}

% unless you've provided the [accepted] option to the mlsys2025


\mlsyssetsymbol{equal}{*}
\mlsyssetsymbol{intern}{$\diamond$}

\begin{mlsysauthorlist}
\mlsysauthor{Yongjun He}{equal,intern,eth}
\mlsysauthor{Shuai Zhang}{equal,aws}
\mlsysauthor{Jiading Gai}{aws}
\mlsysauthor{Xiyuan Zhang}{aws}
\mlsysauthor{Boran Han}{aws}
\mlsysauthor{Bernie Wang}{aws}
\mlsysauthor{Huzefa Rangwala}{aws}
\mlsysauthor{George Karypis}{aws}
\end{mlsysauthorlist}

\mlsysaffiliation{aws}{Amazon Web Services}
\mlsysaffiliation{eth}{ETH Z\"{u}rich}

\mlsyscorrespondingauthor{Yongjun He}{yongjun.he@inf.ethz.ch}
\mlsyscorrespondingauthor{Shuai Zhang}{shuaizs@amazon.com}

\mlsyskeywords{Machine Learning, MLSys}

\vskip 0.3in

\input{0-abstract}
]



\printAffiliationsAndNotice{\mlsysEqualContribution \textsuperscript{$\diamond$}Work done during internship at AWS} % otherwise use the standard text.

\input{1-introduction}
\input{2-background}
\input{3-scheduling}
\input{4-system}
\input{5-evaluation}
\input{6-discussion}
\input{7-related-work}
\input{8-conclusion}
% \input{acknowledgement}


\bibliography{example_paper}
\bibliographystyle{mlsys2025}

\clearpage
\input{appendix}

\end{document}

%% file: 0-abstract.tex
\begin{abstract}
As large language models (LLMs) continue to scale and new GPUs are released even more frequently, there is an increasing demand for LLM post-training in heterogeneous environments to fully leverage underutilized mid-range or previous-generation GPUs and alleviate the shortage of homogeneous high-end GPUs within a single availability zone.
However, achieving high-performance reinforcement learning (RL) training for LLMs on such computing resources remains challenging, as the workflow involves multiple models and tasks with complex computational and data dependencies.
In this paper, we present \sys, a distributed system for efficient RL training in infrastructures with heterogeneous GPUs and networks.
\sys formulates RL training scheduling in heterogeneous
environments as a constrained joint optimization problem
and provides two complementary approaches for addressing this problem: (1) a hybrid scheduling algorithm that efficiently identifies near-optimal solutions, and (2) an integer linear programming (ILP)-based scheduling algorithm that obtains optimal solutions, enabling flexible trade-offs between solution optimality and efficiency.
Our extensive evaluation, consuming 20,000 GPU-hours, shows that \sys achieves up to 9.17$\times$ the throughput of state-of-the-art systems, and 3.17$\times$ on average, across a wide range of workloads and settings.

\end{abstract}

%% file: 1-introduction.tex
\begin{figure*}[t]
\centering
\includegraphics[width=\textwidth]{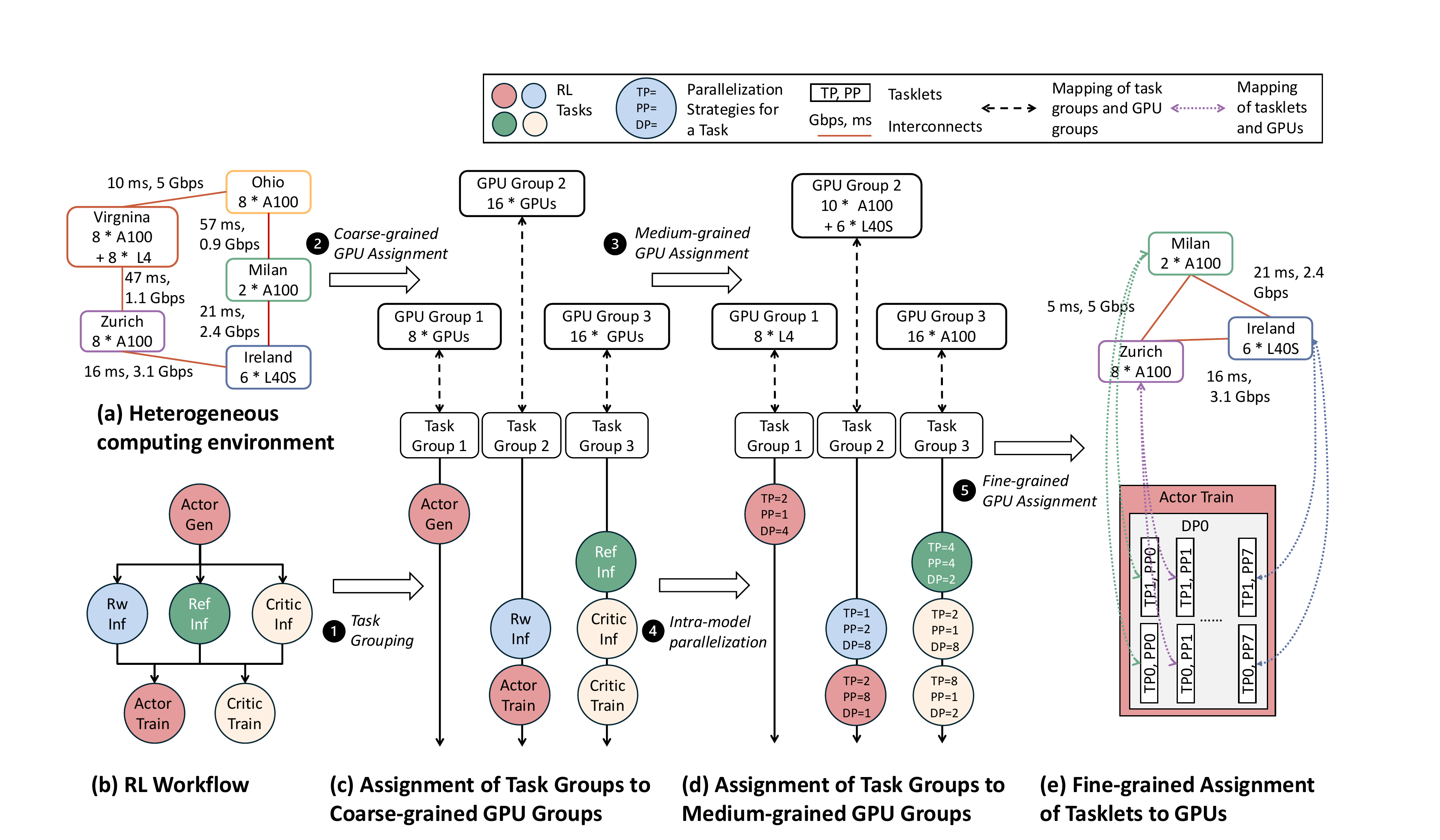}
\caption{
Given a heterogeneous computing environment (a) and an RL workflow (b), \sys employs a multi-level search framework to progressively construct candidate execution plans through \circled{1} task grouping, \circled{2} coarse-grained GPU grouping, \circled{3} medium-grained assignment of task groups to specific GPUs, \circled{4} intra-model parallelization, and \circled{5} fine-grained mapping of tasklets to GPUs.
It enables efficient exploration of near-optimal solutions across sub-problems, while jointly optimizing task/model colocation, parallelization strategies, and device placement under heterogeneous compute and network resources.
}
\label{fig:overview}
\end{figure*}

\section{Introduction}
Reinforcement learning (RL) has become the predominant technique for improving the reasoning ability of large language models (LLMs) and aligning LLMs with human values~\cite{Tulu3,DBLP:journals/tmlr/KaufmannWBH25,DBLP:conf/acl/AhmadianCGFKPUH24,DBLP:journals/tmlr/CasperDSGSRFKLF23}.
Despite the leading performance brought by RL to LLMs, however, comes an explosive growth in computational demand~\cite{Qwen3,DeepSeek-V3,Llama3,LlamaRL}.
In practice, current deployments of RL training relies on individual clusters with a large number of homogeneous GPUs and high-bandwidth networks to meet the computational requirements, as state-of-the-art (SoTA) RL training systems (e.g., verl~\cite{verl} and OpenRLHF~\cite{OpenRLHF}) are tailored for homogeneous computing resources.
On the other hand, as vendors have released an array of new GPU models in recent years, there are a substantial number of mid-range or previous-generation GPUs remaining underutilized across data centers around the world~\cite{DBLP:conf/euromlsys/StratiEKK24,jiangdemystifying,Helix,HeterMoE,DBLP:conf/icse/GaoHLZLLZ0WZGTY24}.
These geo-distributed heterogeneous GPUs collectively provide substantially more memory and compute resources than individual clusters of homogeneous GPUs.
This motivates us to explore an alternative solution by \textit{deploying RL training across a set of heterogeneous GPUs connected via heterogeneous networks}.

Recent studies~\cite{DTFM, Helix, HeterMoE, HexGen, Metis, Sailor, DBLP:conf/euromlsys/StratiEKK24} have investigated the deployment of LLM training and serving in heterogeneous environments to improve the utilization of mid-range or previous-generation GPUs, all centered around the problem of allocating GPU resources to a single model or a single task.
However, the complexity of the RL workflow presents unique obstacles to high-performance deployment in heterogeneous environments, which cannot be addressed by existing methods.
Unlike LLM training and serving, which only involves one model, typical RL workflow~\cite{PPO,GRPO,DPO,SafeRLHF} consists of multiple models and tasks with complex dependencies.
For instance, the most widely used RL algorithm, Proximal Policy Optimization (PPO)~\cite{PPO} (Figure~\ref{fig:overview}(b)), incorporates four LLMs: an \textit{actor} model, a \textit{critic} model, a \textit{reward} model and a \textit{reference} model; and six tasks: \textit{actor generation, reference inference, critic inference, reward inference, actor training}, and \textit{critic training}.
Given the heterogeneity of GPU models and interconnects, the dependencies between different models and tasks, and their different computational characteristics, the desired scheduling algorithm required for efficient RL training needs to jointly optimize (1) the colocation of models and parallelism between tasks (Figure~\ref{fig:overview}(c)); (2) the parallelization of computations within each model and each task (Figure~\ref{fig:overview}(d)); and (3) the fine-grained assignment of tasklets to the heterogeneous devices (Figure~\ref{fig:overview}(e)).

Another category of recent studies has investigated efficient deployment of RL training, but their designs are tailored to clusters with homogeneous GPUs and high-bandwidth networks~\cite{verl,RLHFuse,AReaL,LlamaRL,AsyncFlow,OpenRLHF,DeepSpeed-Chat,NeMo-Aligner}, and do not fully account for heterogeneity in the computing environment within their search space.
An out-of-the-box solution for scheduling RL training in heterogeneous environments is to apply prior heterogeneity-aware scheduling algorithms for LLM training and serving to existing RL training systems.
However, these methods~\cite{DTFM, Helix, HexGen, Metis} focus on the scheduling of a single model/task and require hundreds to thousands of seconds to search for the locally near-optimal execution plan for a single model/task (rather than the entire workflow).
Considering that RL workflows involve multiple models and tasks with complex computational and data dependencies, directly applying prior methods to existing RL training systems is neither practical nor scalable.

To cope with the challenges described above, we propose \sys, a distributed system for efficient RL training over a set of GPUs with heterogeneity in compute and memory resources as well as network interconnects.
Our contributions are summarized as follows:
\begin{itemize}[leftmargin=*,itemsep=-2pt,topsep=-2pt]
    \item We formulate RL training scheduling in heterogeneous environments as a constrained joint optimization problem and introduce a hybrid scheduling algorithm that efficiently identifies near-optimal solutions by (1) decomposing the complex search space with a multi-level search framework (Figure~\ref{fig:overview}); (2) allocating the search budget using the successive halving algorithm~\cite{SHA} (SHA); and (3) generating low-level plans with an evolutionary algorithm (EA).
    We also provide an integer linear programming (ILP)-based scheduling algorithm to obtain optimal solutions, enabling flexible trade-offs between solution optimality and efficiency.
    \item We implemented our proposed scheduling algorithms and built \sys around it on top of verl~\cite{verl}, including a scheduler, a profiler, and an execution engine with extended support for fine-grained resource assignment and load balancing.
    \item We conduct a comprehensive evaluation to compare the performance of \sys and SoTA systems across various workloads and heterogeneous environments.
    The results show that \sys attains throughput up to $9.17\times$ that of SoTA systems, and $3.17\times$ on average.
\end{itemize}

%% file: 2-background.tex
\section{Background and Motivation}
\label{sec:background}

\subsection{Reinforcement Learning for LLMs}
Typical RL workflows~\cite{PPO,GRPO,DPO,SafeRLHF} involve multiple models and tasks with complex computational and data dependencies.
Next, we detail the workflow using PPO~\cite{PPO} (Figure~\ref{fig:overview}(b)) as an example.

\textit{RL Models}.
The \textit{actor} model is the main model to be trained with RL, initialized from pre-trained LLMs.
The \textit{reference} model is a frozen copy of the initial actor model. It’s used to compute the Kullback-Leibler (KL) penalty that ensures the new actor model close to the style and knowledge of the pre-trained model.
The \textit{reward} model is a separate model trained from human preference data. It guides the actor model to generate responses that align with human values and is also frozen during RL training.
The \textit{critic} model evaluates the advantage values of the actions taken by the actor model. Its parameters are also updated during RL training.
Another recently proposed representative RL algorithm, GRPO~\cite{GRPO}, accelerates RL training by eliminating the need for a separate critic model.
The architectures and parameters of the reward and critic models can differ from those of the actor and reference models.

\textit{RL Tasks}.
At the beginning of the RL workflow, \textit{actor generation} uses the prompts of the training dataset as input and generates responses using the actor model.
Next, \textit{reference inference} uses the prompts and generated responses as input and calculate the reference log probability of each token using the reference model.
Using the same set of prompts and generated responses as input, \textit{reward inference} calculate per-sample scores using the reward model, and \textit{critic inference} calculates per-sample values using the critic model.
Finally, \textit{actor training} and \textit{critic training} use the results from the three inference tasks as input to perform forward and backward passes to update the parameters of the actor model and the critic model, respectively.

Recent studies have also investigated asynchronous RL~\cite{AsynchronousRLHF,AReaL,AsyncFlow,StreamRL} to improve GPU utilization for RL training.
Asynchronous RL training overlaps the time-consuming actor generation of the next few iterations with ongoing training to achieve speedup.
However, it leads to decreased accuracy due to data staleness, and increases memory consumption due to the need to maintain two separate copies of the actor model for generation and training.

\subsection{Parallelization in Distributed Deep Learning.}
To distribute deep learning (DL) workloads over computing devices, three primary parallelization strategies have been proposed.
Correctly combining and tuning them (Figure~\ref{fig:overview}(e) bottom) can significantly improve LLM training and serving performance~\cite{Alpa,AlpaServe}.

\textit{Data parallelism (DP)}~\cite{ZeRO}.
Each DP group has a full copy of the model weights and processes a subset of the input dataset, and all DP groups periodically synchronize their model weights using averaged gradients.

\textit{Pipeline parallelism (PP)}~\cite{GPipe}.
The layers of a model are partitioned across PP groups.
By splitting the training batch into multiple micro-batches, the forward and backward passes of different micro-batches can be pipelined across all PP group.

\textit{Tensor parallelism (TP)}~\cite{Megatron-LM}.
The weight matrices of each layer are partitioned across TP groups along the row or column dimension, and all TP groups perform all-reduce to aggregate the output of each partitioned GEMM.

\subsection{Motivation}
The heterogeneous characteristics of today's computing environments and RL workflows reveal opportunities to improve GPU utilization by deploying RL training in heterogeneous environments.
Nevertheless, existing systems exhibit limitations that prevent them from realizing the full potential of such deployments.

\subsubsection{Opportunities}
\textbf{Heterogeneity in computing environments.}
Recent studies~\cite{DBLP:conf/euromlsys/StratiEKK24,jiangdemystifying,Helix,HeterMoE,DBLP:conf/icse/GaoHLZLLZ0WZGTY24} investigates the availabilities of GPU resources, showing that there are severe shortages of homogeneous high-end GPUs within a single availability zone, while substantial heterogeneous GPUs are available across geographical locations.
Therefore, training~\cite{DTFM,HeterMoE,Sailor} and serving~\cite{Helix,ThunderServe,HexGen} LLMs in heterogeneous environments have attracted research interests as it allows resource-intensive LLM jobs to leverage all available GPU resources for acceleration.

\textbf{Heterogeneity in RL workflows.}
In RL workflows, the actor, critic, reference, and reward models may use LLMs with different model sizes and perform generation, inference, or training during different tasks.
Therefore, RL workflows have different requirements for computation, memory, and communication across tasks.
For example, the actor generation is memory bound~\cite{zhang2024} and needs to maintain key-value cache (KV cache)~\cite{vLLM}; and the actor/critic training are computation bound~\cite{LLMStation} and needs to maintain activations, gradients and optimizer states~\cite{ZeRO}.
In comparison with LLM training and serving which involves only single model or single task, the heterogeneous characteristics of the RL workflow make it a workload with new potential for utilizing otherwise idle heterogeneous GPU resources.

\subsubsection{Limitations}
\textbf{Limited search space in existing RL training systems.}
To accelerate RL workflows with heterogeneous characteristics, and complex computational and data dependencies, a flurry of RL training systems~\cite{verl, OpenRLHF, DeepSpeed-Chat, NeMo-Aligner, RLHFuse, LlamaRL, AReaL, AsyncFlow} have been proposed.
However, they are all tailored for homogeneous GPUs with high-bandwidth networks, and almost none of them fully account for the heterogeneity of hardware and networks in their search space.
StreamRL~\cite{StreamRL} is perhaps the most relevant effort to ours, organizing heterogeneous GPUs into two groups: one group for actor generation and a separate group for the remaining tasks.
However, it requires that all GPUs within the same group are homogeneous and located in the same data center.

\textbf{Time-consuming heterogeneity-aware scheduling algorithms.}
A natural approach to enhance RL training systems is to apply the heterogeneity-aware scheduling algorithms originally designed for LLM training and serving to the scheduling of each model and each task in the RL workflow.
As reported by verl~\cite{verl} and RLHFuse~\cite{RLHFuse}, searching for efficient deployment plans for RL training on homogeneous GPUs connected over a homogeneous network requires examining over millions to billions of plans, which takes hundreds to thousands of seconds.
Meanwhile, the heterogeneity-aware scheduling algorithms originally designed for LLM training~\cite{DTFM,Metis} and serving~\cite{Helix,ThunderServe,HexGen} require $1,000{\sim}10,000\times$ more search time to search for the locally near-optimal plan for a single model/task (rather than the entire workflow), so this naive combination is neither practical nor scalable for real-world deployment.

%% file: 3-scheduling.tex
\section{Scheduling in \sys}
\label{sec:scheduling}
We begin by formulating RL training scheduling in heterogeneous environments as a constrained joint optimization problem over partitioning and assignment strategies. 
We then present our multi-level search framework for decomposing the search space, along with a cost model for efficient execution time estimation. 
Finally, we introduce two complementary scheduling algorithms: (1) a hybrid approach based on successive halving and evolutionary search that efficiently identifies near-optimal solutions, and (2) an ILP-based approach that provides optimal solutions when sufficient computational resources are available.

\subsection{Problem Formulation}
\label{subsec:problem}
\textbf{Notation.}
Let $\mathbf{G}^t = (V^t, E^t)$ denote the computational graph of the $t$-th task in an RL workflow, where $t \in \{1,\ldots,T\}$, $V^t$ denotes the set of computational operators, and $E^t$ denotes the set of tensors shared between operators.
The overall computational graph of the RL workflow is then defined as $\mathbf{G} = (\bigcup_{t=1}^T V^t, \bigcup_{t=1}^T E^t \cup E^\mathrm{inter})$, where $E^\mathrm{inter}$ denotes the set of edges between tasks.

Let $\mathbf{G_D} = (V_D, E_D)$ denote the device topology graph for a heterogeneous environment, where $V_D = \{d_1,\ldots,d_N\}$ are a set of $N$ devices and $E_D \subseteq V_D \times V_D$ are communication channels between devices.
Each device $d$ is labeled with computation capability, memory capacity, and HBM bandwidth.
Each edge between $d$ and $d^{\prime}$ is labeled with the latency and bandwidth.

A partitioning strategy $\rho$ transforms the given $\mathbf{G}$ into a new tasklet graph $\mathbf{G_L} = (V_L, E_L)$ by first partitioning the operators with intra-model parallelization and then merging them into tasklets.
Each new node $l_{i,j,k}^t \in V_L$ is a tasklet, $E_L$ denotes the set of tensors transferred between tasklets, and $t,i,j,k$ represent the indices of a tasklet in RL tasks, data parallelism, pipeline parallelism, and tensor parallelism, respectively.
An assignment strategy $\sigma: V_L \to V_D$ assigns, for each tasklet $l \in V_L$, a device $d \in V_D$.
Let $C$ denote a cost model that estimates the execution time per iteration of a given workflow, conditioned on the given resource, partitioning strategy, and assignment strategy.
The functions $M_{\mathrm{working}}(l)$ and $M_{\mathrm{model}}(l)$ represent the working and model memory consumption of tasklet $l$, respectively, while $M_{\mathrm{gpu}}(d)$ denotes the GPU memory capacity of device $d$.

We defer the notation for device and network attributes to Appendix~\ref{app:cost_model}, notation for the PPO definition to Section~\ref{subsec:cost_model}, and notation for the scheduling algorithm to Algorithm~\ref{algo:search}.

\begin{definition}[Heterogeneity-Aware RL Training Scheduling Problem]
    \label{def:hetrl}
    Given a computational graph $\mathbf{G}$ for an RL workflow and a device topology graph $\mathbf{G_D}$ for a heterogeneous environment, the \textit{heterogeneity-aware RL training scheduling problem} is to determine an optimal scheduling strategy, which consists of a partitioning strategy $\rho$ and an assignment strategy $\sigma$, such that the execution time of the RL workflow is minimized and resource constraints are satisfied:

    \vspace{-1.0em}
    \begin{small}
    \begin{align*}
        \min_{\rho, \sigma} \quad & C(\rho, \sigma; \mathbf{G}, \mathbf{G_D}) \\
        \text{s.t.} \quad
        & |\{(i,j,k) : l^t_{i,j,k} \in V_L\}| \leq |V_D|, \forall t \in \{1,\ldots,T\} \tag{C1} \\
        % & \sigma^{-1}(a) \cap \sigma^{-1}(a^\prime) = \emptyset, \forall a \neq a^\prime \in V_A \tag{C2} \\
        & \bigcup_{d \in V_D} \sigma^{-1}(d) = V_L \tag{C2} \\
        & \max_{l \in \sigma^{-1}(d)} M_\mathrm{working}(l) + \sum_{l \in \sigma^{-1}(d)} M_\mathrm{model}(l) \\
        & \qquad \quad \leq M_\mathrm{gpu}(d), \forall d \in V_D \tag{C3} \\
    \end{align*}
    \end{small}
\end{definition}

\begin{proposition}
    \label{pro:np_hardness}
    The abstract heterogeneity-aware RL training scheduling problem defined in Definition~\ref{def:hetrl} is NP-hard.
\end{proposition}
The proof is deferred to Appendix~\ref{app:proof}.

\subsection{Multi-Level Search Framework}
\label{subsec:framework}
\textbf{Design.} Given that the heterogeneity-aware RL training scheduling problem is NP-hard, exhaustively searching over all feasible execution plans is computationally intractable.
We propose a multi-level search framework that decomposes the search space into structured subspaces, enabling termination of  poor-performing high-level decisions and efficient exploration of lower-level decisions.
As illustrated in Figure~\ref{fig:overview}, the framework consists of the following levels:
\begin{itemize}[leftmargin=*,itemsep=-2pt,topsep=-2pt]
    \item Level 1 (Task grouping): Given an RL training pipeline, we first partition tasks into disjoint task groups.
    Tasks within the same task group are executed on a shared set of GPUs, with their associated models co-located.
    \item Level 2 (Coarse-grained GPU assignment): Given the number of task groups, we partition the GPUs into disjoint GPU groups and assign each task group to one GPU group.
    At this step, we determine only the number of GPUs in each group, rather than the specific GPU assignments.
    \item Level 3 (Medium-grained GPU assignment): At this step, we generate candidate assignments that map task groups to specific GPUs.
    \item Level 4 (Intra-model parallelization): Given a set of candidate medium-grained GPU assignments, we determine feasible parallelization strategies for individual tasks within each assignment.
    Applying these strategies further decomposes tasks into finer-grained tasklets.
    \item Level 5 (Fine-grained GPU assignment): Finally, we generate candidate assignments that map tasklets to specific GPUs, yielding complete execution plans.
\end{itemize}
Existing RL training frameworks already expose several configurable dimensions that naturally correspond to different levels of the search space.
In particular, task and model colocation (Level 1), placement groups and resource pools (Level 2), and parallelization strategies (Level 4) are commonly supported configuration knobs.
To fully account for heterogeneity in the computing environment within the search space, we extend these dimensions with Level 2 and Level 5, which enables fine-grained tasklet-to-device mapping beyond what is typically supported in current systems.

\textbf{Interpretation.}
Our framework can be viewed as a coarse-to-fine constructive approach that operationalizes the joint optimization over $(\rho,\sigma)$ defined in Section~\ref{subsec:problem}.
Concretely, Levels 1 and 4 instantiate the partitioning strategy $\rho$, performing task grouping and intra-model parallelization to produce the tasklet graph $G_L$, while Levels 2, 3, and 5 instantiate the assignment strategy $\sigma$, generating coarse-to-fine GPU assignments that map computational subgraphs (or nodes) to devices.

\textbf{Search space analysis.}
The search space induced by the multi-level search framework can be characterized as a hierarchical composition of combinatorial subspaces across five levels.
We analyze the search space in a level-wise manner:
\begin{itemize}[leftmargin=*,itemsep=-2pt,topsep=-2pt]
    \item Level 1: The search space of this level for $T$ tasks has size $B_T$, corresponding to the number of set partitions, where $B_T$ denotes the $T$-th Bell number~\cite{rota1964number}. 
    \item Level 2: For simplicity, we consider the case where each task forms its own GPU group, which yields a worst-case upper bound.
    The search space of this level for $N$ GPUs has size $\binom{N-1}{T-1}$, corresponding to the number of integer partitions of $N$ into exactly T positive parts.
    \item Level 3: Given a coarse-grained GPU assignment with GPU group sizes $\{n_t\}_{t=1}^{T}$, where $\sum_{t=1}^{T} n_t = N$, the search space of this level has size $\frac{N!}{\prod_{t=1}^{T} n_t!}$, corresponding to the multinomial coefficient.
    \item Level 4: For simplicity, we assume that each of data, tensor, and pipeline parallelism has a uniform degree.
    Therefore, the search space of this level is upper bounded by
    $\prod_{t=1}^{T} |\{ (i,j,k) \in \mathbb{N}_+^3 : i \cdot j \cdot k \le n_t \}|$.
    \item Level 5: Given the parallelization strategies, each task is further decomposed into tasklets.
    Let $|V_L^t|$ denote the number of tasklets assigned to task group $t$.
    The search space of this level is upper bounded by $\prod_{t=1}^T n_t^{|V_L^t|}$.
\end{itemize}
Note that the search space at each level listed above is conditional on the decisions made in preceding levels, rather than independent.
Combining the above levels, the overall search space is upper bounded by

\vspace{-1.2em}
\begin{scriptsize}
    \begin{equation*}
        B_T \cdot \binom{N-1}{T-1} \cdot \frac{N!}{\prod_{t=1}^{T} n_t!} \cdot \prod_{t=1}^{T} |\{ (i,j,k) \in \mathbb{N}_+^3 : i \cdot j \cdot k \le n_t \}| \cdot \prod_{t=1}^T n_t^{|V_L^t|} .
    \end{equation*}
\end{scriptsize}

\subsection{Cost Model}
\label{subsec:cost_model}
Because the search space is large and running RL training in practice can take tens of minutes per step, we develop a cost model within our framework to quickly estimate execution time.
Cost models differ across RL algorithms and between synchronous and asynchronous modes.
To instantiate the cost model, we formalize a synchronous variant of PPO and connect it to the computational graph defined in Section~\ref{subsec:problem}.

\textbf{PPO.}
We consider a PPO setting with a policy (actor) model $\pi_\theta$, a value function (critic model) $v_\phi$, a reward model $r_\psi$, and a fixed reference policy (model) $\pi_{\mathrm{ref}}$.
Given queries $x$ sampled from the training dataset $\mathcal{D}$ and responses $y$ generated by the old policy model $\pi_{\theta_\mathrm{old}}$, the reward at the $\tau$-th generation step is defined as:
\begin{equation*}
    r_\tau = r_\psi(x, y_{\leq \tau}) - \beta \log \frac{\pi_\theta(y_\tau \mid x, y_{< \tau})}{\pi_{\mathrm{ref}}(y_\tau \mid x, y_{< \tau})},
\end{equation*}
where $\theta$, $\phi$, and $\psi$ denote the parameters of the actor, critic, and reward models, respectively, and $\beta$ is the coefficient of the KL penalty.
Based on the rewards and the learned value function, the estimated advantage $\hat{A}_\tau$ is then computed using Generalized Advantage Estimation (GAE)~\cite{GAE}.
Finally, PPO updates the policy model via a clipped surrogate objective:

\vspace{-1.5em}
\begin{scriptsize}
\begin{align*}
\mathcal{J}_{\mathrm{PPO}}(\theta)
&= \mathbb{E}_{x \sim \mathcal{D}, y \sim \pi_{\theta_\mathrm{old} (\cdot \mid x)}} \big[
\frac{1}{|y|} \sum_1^{|y|} \min \big( \frac{\pi_\theta(y_\tau \mid x, y_{< \tau}))}{\pi_{\theta_\mathrm{old}}(y_\tau \mid x, y_{< \tau}))} \hat{A}_\tau, \\
& \mathrm{clip} \big(
\frac{\pi_\theta(y_\tau \mid x, y_{< \tau}))}{\pi_{\theta_\mathrm{old}}(y_\tau \mid x, y_{< \tau}))} , 1 - \epsilon, 1 + \epsilon \big) \hat{A}_\tau \big) \big]
\end{align*}    
\end{scriptsize}
where $\epsilon$ controls the clipping range,
and updates the value function via a mean-squared error loss.

From the above formulation of PPO, we observe that the training process can be decomposed into multiple computation stages with well-defined dependencies.
Additionally, a synchronous variant of PPO enforces an iteration-level barrier: the actor generation of the next iteration cannot begin until the training of the current iteration completes.
We therefore model synchronous PPO as a computational graph $\mathbf{G}$ formed by a set of task-level computational graphs $\{\mathbf{G}^t\}_{t=1}^6$, where actor generation ($\mathbf{G}^1$) is followed by reward, reference, and critic inference ($\mathbf{G}^2$--$\mathbf{G}^4$) in parallel, before actor and critic training ($\mathbf{G}^5$ and $\mathbf{G}^6$).

\textbf{Cost model for synchronous PPO.}
We adopt a compact convention and overload the notation: $C$ denotes both the cost model and its evaluated cost under the current inputs.
We now present a concrete instantiation of $C$ for PPO:
\begin{equation*}
    C_\mathrm{SyncPPO} = C^1 + \Phi(\{C^2, C^3, C^4\}) + \Phi(\{C^5, C^6\}),
\end{equation*}    
where $C^t$ denotes the cost associated with the $t$-th task ($\mathbf{G}^t$) for $t \in \{1,\dots,6\}$, corresponding to actor generation, reward inference, reference inference, critic inference, actor training, and critic training, respectively.
$\Phi$ aggregates the costs of tasks without dependencies and is defined as:
\begin{equation*}
    \Phi(\{C^t\}) = \eta \max_t C^t + (1 - \eta)\sum_t C^t, \eta \in [0,1],
\end{equation*}
where the coefficient $\eta$ parameterizes the level of task parallelism (0: sequential; 1: fully parallel; else: partial).
The cost model of the $t$-th task is given by:

\vspace{-1.0em}
\begin{small}
\begin{equation*}
C^t =
\begin{cases}
    \Psi^\mathrm{gen}(C_\mathrm{comp}^t, C_\mathrm{hbm}^t, C_\mathrm{tp}^t, C_\mathrm{pp}^t), & t = 1,\\
    \Psi^\mathrm{inf}(C_\mathrm{comp}^t, C_\mathrm{tp}^t, C_\mathrm{pp}^t), & t \in \{2,3,4\},\\
    \Psi^\mathrm{train}(C_\mathrm{comp}^t, C_\mathrm{tp}^t, C_\mathrm{pp}^t, C_\mathrm{dp}^t, C_\mathrm{bubble}^t), & t \in \{5,6\}, \\
\end{cases}
\end{equation*}    
\end{small}
where $C_\mathrm{comp}^t$ estimates a set of per-tasklet or per-subgraph computation costs, $C_\mathrm{hbm}^t$ estimates a set of per-tasklet or per-subgraph HBM costs (i.e., loading LLMs from HBM to SRAM during decoding), $C_\mathrm{bubble}^t$ estimate a set of per-tasklet or per-subgraph pipeline-bubble costs, and $C_\mathrm{tp}^t$, $C_\mathrm{pp}^t$, and $C_\mathrm{dp}^t$ estimate a set of per-tasklet or per-subgraph TP, PP, and DP computation costs.
$\Psi^\mathrm{gen}$, $\Psi^\mathrm{inf}$, and $\Psi^\mathrm{train}$ estimate task-level costs based on the given sets of tasklet- or subgraph-level costs.
%\textcolor{blue}{, distinct from the parameter $\psi$ used for the reward model}
For brevity, we have omitted some inputs to the above equations, including device and network attributes, and also omitted the resharding cost in the end-to-end cost.
Details can be found in Appendix~\ref{app:cost_model}.

\input{algo-search}

\subsection{Hybrid Scheduling Algorithm}
\textbf{SHA for search budget allocation.}
Although theoretically identifying the optimal high-level decision is intractable due to the NP-hardness of the problem, we aim to prioritize promising high-level decisions in a principled manner while keeping the method scalable and simple from an engineering perspective.
To this end, we adopt SHA~\cite{SHA} for search budget allocation, a robust and straightforward bandit-based algorithm designed to the non-stochastic best arm identification problem within a fixed budget.
Next, we determine at which levels to apply SHA.
In the bandit formulation, each high-level decision corresponds to an arm.
Applying SHA only to Level 1 results in too few arms, yielding limited speedup, as the number of elimination rounds is bounded by $\lceil \log_2 B_T \rceil$.
In contrast, applying SHA to Levels 1–3 introduces an excessive number of arms due to the combinatorial explosion of the search space, making it infeasible to evaluate them within a fixed budget.
Therefore, we apply SHA to Levels 1 and 2, which provide a balanced number of arms for effective speedups while keeping the budget allocation tractable.

The pseudocode in Algorithm~\ref{algo:search} shows how \sys extends SHA to a nested form and applies it to our problem.
\sys treats task groupings as arms at Level 1 and GPU groupings as arms at Level 2 in the multi-armed bandits problem; the execution time estimated by the cost model serves as each arm's loss.
Concretely, the inputs are the user-defined budget $B$, workflow information and device information.
In our problem, the search budget corresponds to the wall-clock time allocated for the scheduling procedure.
At Level 1 (lines 14-16 and lines 29-33), it first assigns a starting budget $b_m$ to each task grouping, which is shared by all the GPU groupings corresponding to this task grouping.
At Level 2 (lines 17-20 and lines 27-29), it also first assigns a starting budget $b_{m,n}$ to each pair of tasking grouping and GPU grouping.
It then uses an evolutionary algorithm at lower levels and a cost model to generate and evaluate $b_{m,n}$ candidate plans (lines 21–25).
Finally, it discards the worst half GPU groupings and continues the procedure with the better half with a doubled budget until the assigned budget $b_m$ is exhausted.
This procedure is also repeated at Level 1 by discarding the worse half tasking groupings and doubling the budget for the next round until the global budget $B$ is exhausted.
At each new Level 1 round, we retain the best half of GPU groupings for each task grouping.
With nested SHA, we allocate more search budget to sets of candidates associated with more promising high-level decisions, while discarding less promising candidates early.
SHA enjoys theoretical guarantees on the probability of selecting an optimal (or near-optimal) configuration within a given budget~\cite{SHA, Hyperband}.

\textbf{Evolutionary algorithm for low-level plan generation.}
Given the task grouping from Level 1 and the coarse-grained GPU assignment from Level 2, we generate fine-grained GPU assignments through Level 3 to 5.
By treating the devices assigned to tasklets within the same pipeline stage as a graph partition, and those assigned to the same task group as a coarsened graph partition, the procedure can be viewed as a graph partitioning problem with a complex objective on the device topology graph.
Following the line of research that uses Evolutionary Algorithm (EA)~\cite{DBLP:journals/tc/BuiM96,DBLP:journals/jgo/SoperWC04,DTFM} for graph partitioning, we develop a EA for low-level plan generation.

Concretely, we randomly initialize medium-grained GPU assignments at Level 3, enumerate all feasible intra-model parallelization strategies at Level 4, and then randomly initialize fine-grained GPU assignments at Level 5.
These fine-grained GPU assignments serve as the initial population of the EA, which produces the next generation as follows.
It first generates a new “offspring” $o$ via mutation from the population.
We than conduct swapping on $o$ to find a new individual $o^*$ that leads to lower cost.
Finally, we add $o^*$ to the population and remove the worst individual in the population if $o^*$ achieves a lower cost.

As described in Section~\ref{subsec:framework}, the multi-level search framework consists of five levels.
While omitting Level 3 does not affect the completeness of the search space, it is introduced to provide a group-level structure that enables GPU mutation and swapping across task/GPU groups, thereby improving the efficiency of the search process.
We customize the mutation operator and incorporate a swap-based local search.
The mutation operator, with a certain probability, replaces a GPU in a training-task group with a higher-TFLOPS one selected from GPUs not assigned to any training-task group.
The local search greedily improves GPU-group locality while keeping group sizes fixed.
For each candidate, we repeatedly evaluate cross-group GPU swaps and apply the one with the largest positive gain in a locality score that captures, but is not limited to, machine-, zone-, and region-level affinities.
The process terminates when no improving swap exists, and the improvements obtained by the phenotype are not mapped back to the genotype or incorporated into the population.
This design allows faster evolution while preserves population diversity and prevents premature convergence~\cite{DBLP:journals/compsys/HintonN87, baldwin1896new}.

\subsection{ILP-Based Scheduling Algorithm}
While the hybrid scheduling algorithm can efficiently identify near-optimal solutions in large-scale deployments, we further provide an ILP-based scheduling algorithm to obtain optimal solutions when sufficient search budgets are available.
Our problem formulation (Section~\ref{subsec:problem}) can be straightforwardly converted into a corresponding ILP formulation by converting the discrete scheduling choices into binary decision variables.
The ILP formulation preserves the same search space as our hybrid scheduling algorithm, but replaces heuristic exploration with exact optimization.

Specifically, for each RL task, we enumerate all feasible parallelization strategies over candidate tensor, data, and pipeline parallelism, and associate each strategy with a binary decision variable.
Based on the selected strategy, we further introduce binary variables to determine whether each tasklet is instantiated and to assign each instantiated tasklet to a specific GPU.
On top of these decision variables, we use the analytical cost model to parameterize the execution cost of each task under heterogeneous devices.
In particular, the formulation captures the memory footprint and execution time of each tasklet.
These tasklet-level costs are then aggregated into task-level durations according to the selected partitioning and assignment strategies.
To model end-to-end execution, we additionally introduce time variables for each task, including start time, duration, and completion time, and minimize the overall workflow makespan.
The formulation not only enforces the constraints defined in Section~\ref{subsec:problem} but also incorporates the task dependencies as constraints.

%First, each task must select exactly one feasible parallelization strategy, and the number of instantiated tasklets must be consistent with the selected strategy.
%Second, each instantiated tasklet must be mapped to exactly one GPU, while the task-level and tasklet-level assignment variables remain consistent.
%Third, the formulation enforces GPU memory-capacity constraints by ensuring that, for each GPU, the active memory footprint of the currently executing task together with the inactive memory footprints of the remaining colocated tasks does not exceed device capacity. Fourth, workflow precedence constraints derived from the RL execution DAG ensure that dependent tasks are executed in the correct order. Finally, resource exclusivity is enforced by preventing overlapping execution of tasks assigned to the same GPU.

%% file: algo-search.tex
\begin{algorithm}[t]
\caption{\sys Hybrid Scheduling Algorithm}
\begin{scriptsize}
\textbf{Input:}
search budget $B$, a computational graph for an RL workflow $\mathbf{G}$, a device topology graph $\mathbf{G_D}$, number of GPUs $N$

\textbf{Output:} Execution plan with the lowest estimated cost
\begin{algorithmic}[1]
\STATE let $TG = \{ tg_{\xi_i} \}$ be the set of all feasible task groupings for a given RL workflow, where $tg_{\xi_i}$ denotes a feasible grouping of RL tasks
\STATE let $GG = \{ tg_{\xi_i} \mapsto GG_{\xi_i} \}$ be the set of all feasible GPU groupings, where $GG_{\xi_i} = \{ gg_{\xi_{i,j}} \}$ denotes the set of all feasible GPU groupings for a given $tg_{\xi_i}$ and $gg_{\xi_{i,j}}$ denotes a feasible grouping of GPUs
\STATE let $C_{\mathrm{plans}} = \{ tg_{\xi_i}, gg_{\xi_{i,j}} \mapsto \{ c_{\xi_{i,j,k}} \} \}$ be the cost of candidate plans, where $c_{\xi_{i,j,k}}$ denote the cost of the $k$-th candidate plans under $tg_{\xi_i}$ and $gg_{\xi_{i,j}}$
\STATE initialize $TG \gets \emptyset, GG \gets \emptyset, C \gets \emptyset$
\STATE $TG \gets$ $\mathrm{TaskGrouping}(\mathbf{G})$
\FORALL{$tg_{\xi_i} \in TG$}
    \STATE $GG_{\xi_i} \gets \mathrm{GPUGrouping}(N, |tg_{\xi_i}|)$
    \STATE $GG \gets GG \cup \{ tg_{\xi_i} \mapsto GG_{\xi_i} \}$
    \FORALL{$gg_{\xi_{i,j}} \in GG_{\xi_i}$}
        \STATE $C_{\mathrm{plans}} \gets C_{\mathrm{plans}} \cup \{ tg_{\xi_i}, gg_{\xi_{i,j}} \mapsto \{ \infty \} \}$
    \ENDFOR
\ENDFOR

\STATE $(TG_0, GG_0) \gets (TG, GG)$
\FOR{$m = 0, 1, \dots, \lceil \log_2(|TG|) \rceil - 1$}
    \FORALL{$tg_{\xi_i} \in TG_m$}
        \STATE let $b_m = \lfloor \frac{B}{|TG_m|\lceil \log_2(|TG|) \rceil} \rfloor$ be search budget for $tg_{\xi_i}$
        \STATE $GG_{\xi_i, 0} \gets GG_m(tg_{\xi_i})$
        \FOR{$n = 0, 1, \dots, \lceil \log_2(|GG_m(tg_{\xi_i})|) \rceil - 1$}
            \FORALL{$gg_{\xi_{i,j}} \in GG_{\xi_i, n}$}
                \STATE let $b_{m, n} = \lfloor \frac{b_m}{|GG_{\xi_i, n}|\lceil \log_2(|GG_m(tg_{\xi_i})|) \rceil} \rfloor$ be search budget for $gg_{\xi_{i,j}}$
                \FOR{$k = 0, 1, \dots, b_{m, n} - 1$}
                    \STATE let $p_k \gets \mathrm{EA}(tg_{\xi_i},gg_{\xi_{i,j}},\mathbf{G_D},\ldots)$ be a new low-level plan
                    \STATE $c_{\xi_{i,j,k}}=C(p_k,\mathbf{G_D},\dots)$
                    \STATE $C_{\mathrm{plans}}(tg_{\xi_i}, gg_{\xi_{i,j}}) \gets C_{\mathrm{plans}}(tg_{\xi_i}, gg_{\xi_{i,j}}) \cup c_{\xi_{i,j,k}}$
                \ENDFOR
            \ENDFOR
            \STATE $GG_{\xi_i, n+1} \gets \mathrm{BestHalf}(GG_{\xi_i, n}, C_{\mathrm{plans}})$
        \ENDFOR
        \STATE $GG_{m+1}(tg_{\xi_i}) \gets \mathrm{BestHalf}(GG_m(tg_{\xi_i}), C_{\mathrm{plans}})$
    \ENDFOR
    \STATE $TG_{m+1} \gets \mathrm{BestHalf}(TG_{m}, C_{\mathrm{plans}})$
\ENDFOR
\label{algo:search}
\end{algorithmic}
\end{scriptsize}
\end{algorithm}

%% file: 4-system.tex
\section{\sys system}
\label{sec:system}
\begin{figure}[t]
\centering
\includegraphics[width=\columnwidth]{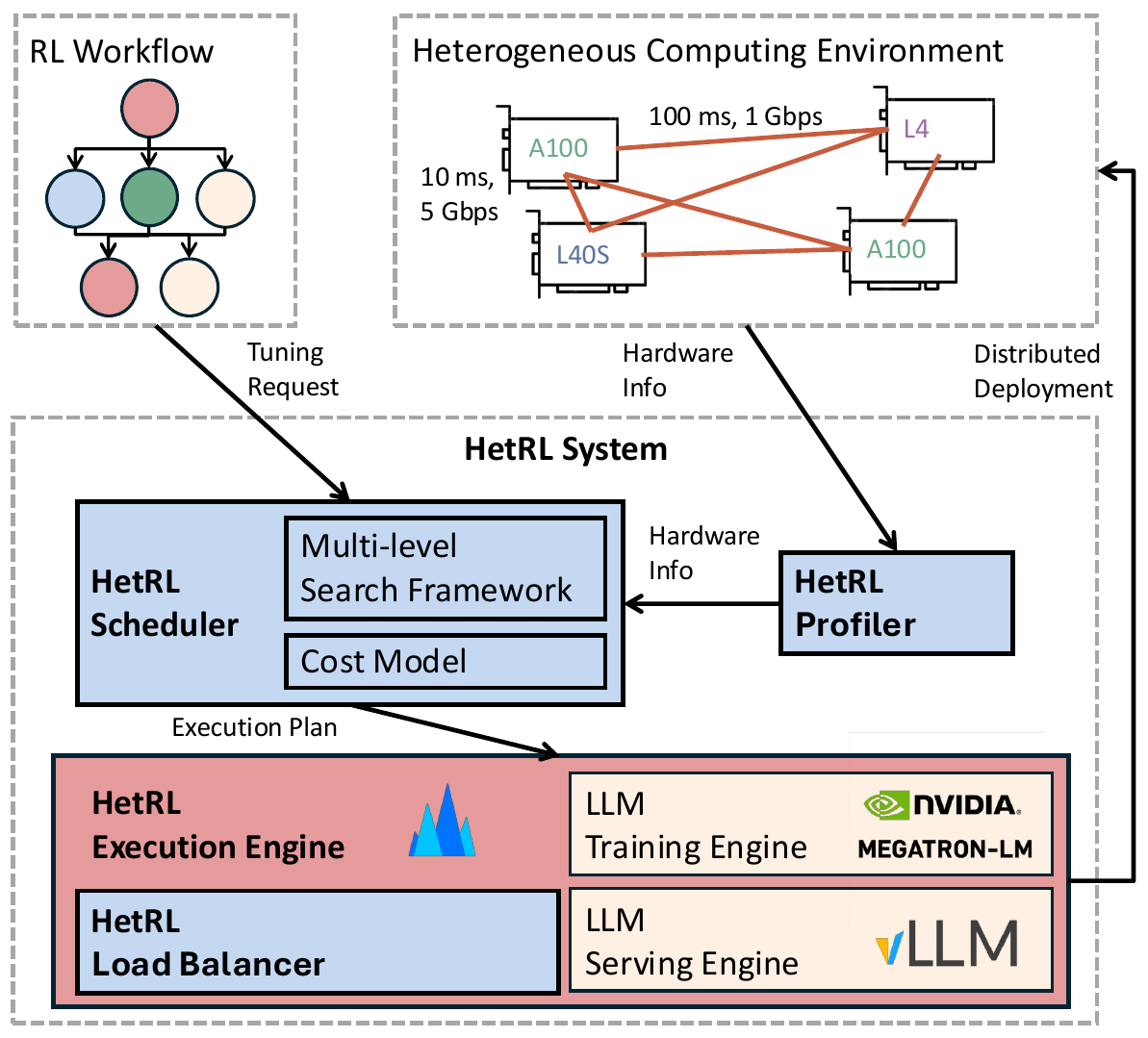}
\caption{\sys system overview.}
\label{fig:system}
\end{figure}

Centering on our proposed scheduling algorithm, we built \sys, a distributed system for efficient RL training in heterogeneous environments.
Figure~\ref{fig:system} illustrates the system overview of \sys, which comprises four key components: a scheduler, a profiler, a load balancer, and a distributed execution engine.
Next, we describe an overview of the system's end-to-end execution and the role of each component.

\subsection{Execution Overview}
An RL training job submitted to \sys contains an RL algorithm, a dataset, models for different tasks, an optimizer, numerical precision, global batch size, sequence lengths of prompts and responses, and other optional configurations.
After receiving a job request, the \sys profiler first collects hardware information about the computing environment, including the computation power (TFLOPs), memory capacity (GBs), and HBM bandwidth (GB/s) of available GPUs, intra-machine bandwidth (GB/s), and network delay (ms) and bandwidth (Gbps) between them.
Then the \sys scheduler searches for a near-optimal execution plan according to the workflow information and hardware information.
As described in Section~\ref{sec:scheduling}, the scheduler leverages the multi-level search framework and our proposed scheduling algorithms for generation and selection of candidate plans, and uses its cost model to evaluate various candidate plans in terms of training throughput and memory footprint.
Finally, the \sys execution engine deploys the RL training job in the heterogeneous environment according to the selected execution plan.
The \sys execution engine is implemented on top of verl~\cite{verl} with extended support for fine-grained resource assignment and load balancing, and it uses Megatron-LM~\cite{Megatron-LM} and vLLM~\cite{vLLM} as training and serving engines.

\subsection{Load Balancing}
The \sys load balancer incorporates several load balancing strategies to better accommodate RL workflows to heterogeneous computing environments.
Load balancing can be roughly divided into two categories: data-level and layer-level.
To achieve data-level load balancing, it adjusts the local batch sizes across GPUs within a DP group during the actor rollout task based on estimates from the cost model.
For the other tasks, where sequence lengths are known beforehand, it assign samples with longer sequence length to more powerful GPUs.
To achieve layer-level load balancing, it adjusts the layer distribution across pipeline stages based on estimates from the cost model.
These three load balancing strategies enlarge the search space of our scheduling algorithms and can be implemented without invasive modifications to mainstream frameworks (e.g., verl, Megatron-LM and vLLM).
We leave the integration of more advanced load balancing strategies~\cite{Metis} as future work.

%% file: 5-evaluation.tex
\section{Evaluation}
\label{sec:evaluation}
We first compare the end-to-end performance of \sys against SoTA RL training systems under various workloads and computing environments.
Subsequently, we evaluate the search efficiency and the effectiveness of load balancing.

\subsection{Experimental Setup}
\label{subsec:setup}
\textbf{Hardware.}
We conduct experiments on a testbed with a total of 64 GPUs, of which 24 are A100s, 24 are L40Ss, and 16 are L4s.
The detail GPU specifications are listed in Table~\ref{tab:gpu}.
To simulate the heterogeneous network environments, we measure the latencies and bandwidths between 10 different regions (Virginia, Ohio, Paris, Stockholm, London, Ireland, Spain, Zurich, Frankfurt, and Milan) and apply them to our testbed, as shown in Figure~\ref{fig:e2e} (a) and (b).
Our evaluation covers four different network environments:
\begin{itemize}[leftmargin=*,itemsep=-2pt,topsep=-2pt]
    \item \textit{Scenario 1 (Single-Region)}. This is a standard setting provided by cloud service providers.
    We do not enforce latency or bandwidth controls here.
    \item \textit{Scenario 2 (Multi-Region-Hybrid)}. We consider individual GPUs across Ohio and Virginia, but subset of Virginia GPUs are at the edge and only have direct connections to GPUs within Virginia.
    The inter-region connections between Ohio and Virginia have 10 ms of delay and 5 Gbps of bandwidth, while connections involving edge GPUs have a bandwidth of 1 Gbps.
    \item \textit{Scenario 3 (Multi-Country)}. We consider individual GPUs cross eight different regions in Europe.
    For inter-region connections, the delay is $5{\sim}30$ ms and the bandwidth is $1.9{\sim}5.0$ Gbps.
    \item \textit{Scenario 3 (Multi-Continent)}. We consider individual GPUs cross eight different regions acroos Europe and US.
    For inter-region connections, the delay is $5{\sim}60$ ms and the bandwidth is $0.9{\sim}5.0$ Gbps.
\end{itemize}

\begin{table}[t]
  \caption{GPU specifications.}
  \centering
  \begin{tabular}{p{0.09\linewidth}
                  p{0.1\linewidth}
                  p{0.09\linewidth}
                  p{0.2\linewidth}
                  p{0.1\linewidth}
                  p{0.1\linewidth}
                  }
    \toprule
    \textbf{Model} & \textbf{Arch}  & \textbf{Size (GB)}  & \textbf{FP16 Perf. (TFLOPS)} & \textbf{HBM (GB/s)} & \textbf{Intra (GB/s)} \\
    \midrule
    A100 & Ampere & 40 & 312 & 2039 & 600 \\
    L40S & Ada & 48 & 366 & 864 & 64 \\
    L4 & Ada & 24 & 121 & 300 & 64 \\
    \bottomrule
  \end{tabular}
  \label{tab:gpu}
\end{table}

\textbf{Models and RL algorithms.}
We choose representative open-source LLMs, the Qwen series, and consider three different sizes of Qwen models, 4B, 8B, and 14B.
For the RL algorithm, we evaluate two popular methods, PPO~\cite{PPO} and GRPO~\cite{GRPO}, considering both their synchronous and asynchronous versions.
In the evaluation, we use the same size of LLM for all models and tasks within each RL workflow.
However, \sys can flexibly accommodate RL workflows that use models of different sizes to perform different tasks.

\textbf{Datasets and hyperparameters.}
We conduct experiments on GSM8k dataset from OpenAI, an open-source, gold-standard benchmark for mathematical reasoning.
We set the maximum length of both input prompts and output responses to 1024, the global batch size to 384, and the number of responses generated per prompt to 8.
In the RL workflow, training uses mixed precision with the Adam optimizer, and inference and generation use BF16 precision.
The above settings are consistent with previous studies on RL training systems~\cite{verl}.

\textbf{Baselines.}
We compare \sys against two baselines:
\begin{itemize}[leftmargin=*,itemsep=-2pt,topsep=-2pt]
    \item verl~\cite{verl} is a widely-used, open-source, SoTA RL training system in homogeneous setting.
    It proposes a hierarchical hybrid programming model to flexibly support various RL workflows and configurations.
    \item StreamRL~\cite{StreamRL} is the SoTA RL training system in heterogeneous and asynchronous setting.
    It supports the deployment of actor generation and the rest tasks in two separate data centers.
\end{itemize}
Since StreamRL is not open-source, we implemented its asynchronous version on top of verl.
In addition, we chose vLLM as the inference engine and Megatron-LM as the training engine for all variants to ensure a fair comparison.
We recognize other well-known RL training systems; however, we do not include them in the evaluation since their features and functionalities largely overlap with the baselines or they focus on RL algorithmic innovations.

Unless otherwise specified, all experiments are conducted under the experimental setup described above.
For search efficiency comparisons, we use different hardware settings (i.e., CPU-only machines), while for training quality evaluations, we use alternative models and datasets, as detailed in the corresponding sections.

\subsection{End-to-End Comparison}
\begin{figure*}[t]
\centering
\includegraphics[width=\textwidth]{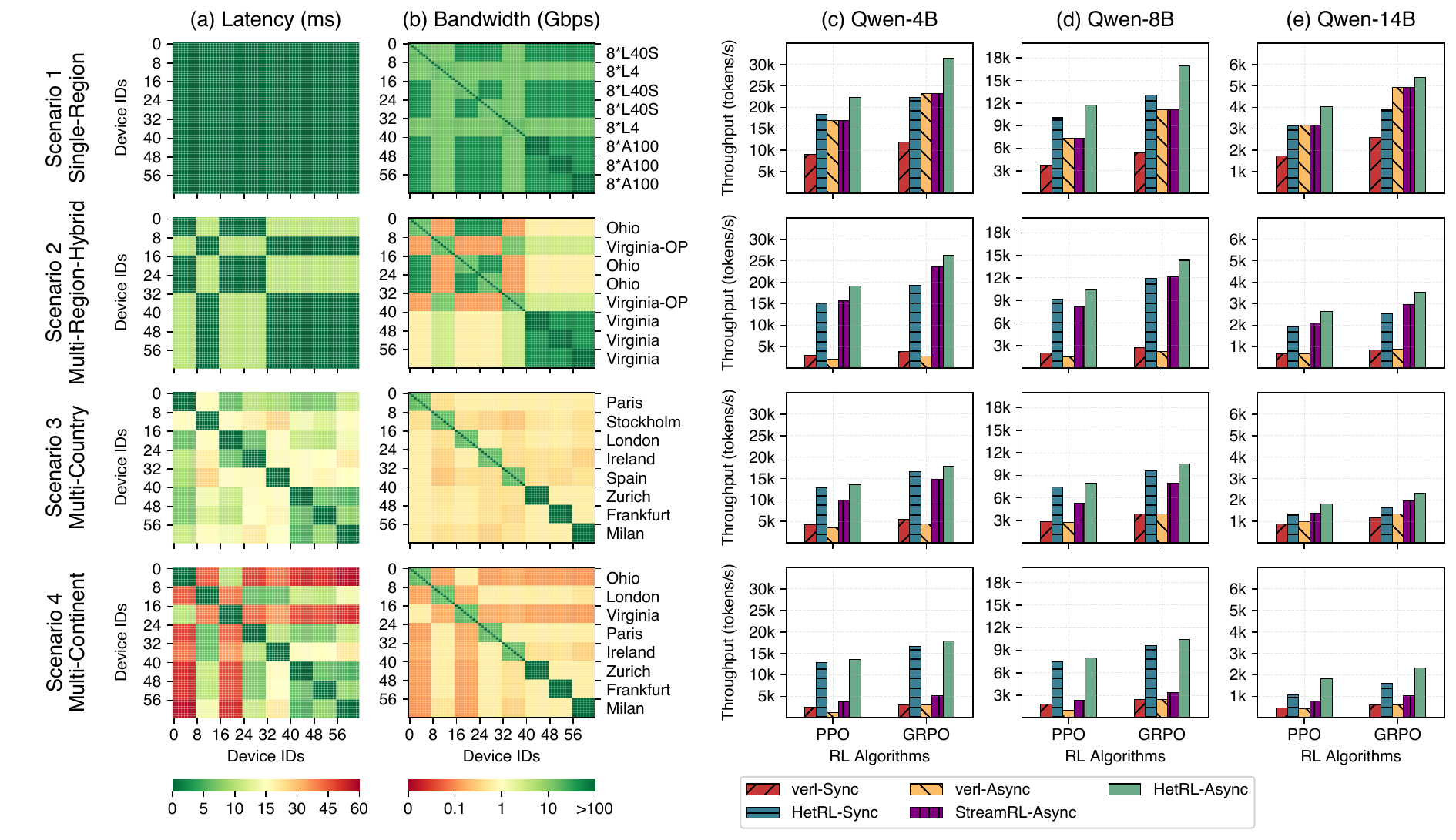}
\caption{End to end compassion of \sys with verl and StreamRL in four different scenarios.
Column (a) and (b) visualize the delay and bandwidth of four scenario respectively;
Column (c), (d), and (e) illustrate the PPO and GRPO throughput comparison respectively.}
\label{fig:e2e}
\end{figure*}

Our first experiment compares end to end performance of \sys against verl and StreamRL in four different scenarios, corresponding to four rows of Figure~\ref{fig:e2e}.
Figure~\ref{fig:e2e}(c - e) show the throughput of training Qwen-4B, 8B, and 14B with PPO and GRPO, respectively.
In Single-Region scenario, \sys outperforms verl by $1.51{\sim}2.05\times$ in synchronous RL training and outperforms StreamRL and verl by $1.1{\sim}1.31\times$ in asynchronous RL training.
In Multi-Region-Hybrid scenario, \sys outperforms verl by $3.01{\sim}4.99\times$ in synchronous RL training and outperforms StreamRL and verl by $1.11{\sim}1.27\times$ and $4.07{\sim}9.17\times$ in asynchronous RL training.
In Multi-Country scenario, \sys outperforms verl by $1.4{\sim}3.07\times$ in synchronous RL training and outperforms StreamRL and verl by $1.19{\sim}1.5\times$ and $1.71{\sim}4.0\times$ in asynchronous RL training.
In Multi-Continent scenario, \sys outperforms verl by $2.24{\sim}5.46\times$ in synchronous RL training and outperforms StreamRL and verl by $2.25{\sim}3.72\times$ and $4.38{\sim}10.76\times$ in asynchronous RL training.

\sys achieves higher training throughputs than baselines in all scenarios by fully considering the characteristics of heterogeneous GPUs and networks in its scheduling algorithm.
The performance gaps in the scenarios 2-4 are larger due to the larger differences in network latencies and bandwidths between devices.
While \sys-Async is always faster than \sys-Sync, we observe that verl-Async is sometimes slower than verl-Sync due to its unoptimized scheduling in heterogeneous environments.
Except for the first scenario, StreamRL-Async achieves higher training throughput than verl-Async because its scheduling algorithm allows dividing GPU resources across data centers into two groups and assigning them to actor generation and the other tasks accordingly.
The performance gaps between \sys and the baselines for PPO and GRPO are different, mainly because GRPO does not have a critic model and critic inference and training tasks.

\subsection{Effectiveness of Load Balancing}
\begin{figure}[t]
\centering
\includegraphics[width=\columnwidth]{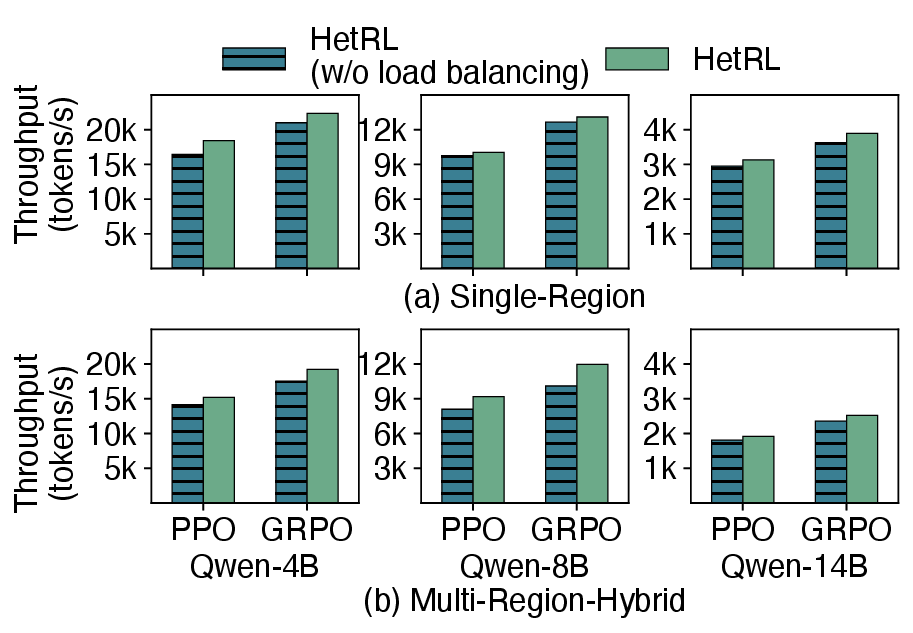}
\caption{Effects of load balancing on synchronous RL training across model sizes under Single- and Multi-Region scenarios.}
\label{fig:load_balancing}
\end{figure}

This section conducts an ablation study on the effect of load balancing.
We evaluate the synchronous RL training of Qwen 4B, 8B, and 14B using PPO and GRPO.
As illustrated in Figure~\ref{fig:load_balancing}, load balancing improves training throughput by up to 12\% in the Single-Region scenario and up to 18\% in the Cross-Region scenario, which confirms the effectiveness of our load balancing strategies.
Our performance gains from load balancing is slightly smaller than related work such as Metis~\cite{Metis}, which is $19{\sim}22\%$.
The gap can be minimized by integrating more proposed load balancing strategies into \sys.
In asynchronous RL training, further improvements are not significant because the resource scheduling between generation and training determines the performance.

\subsection{Effectiveness of the Scheduling Algorithm}
\label{subsec:search_exp}

\begin{figure}[t]
\centering
\includegraphics[width=\columnwidth]{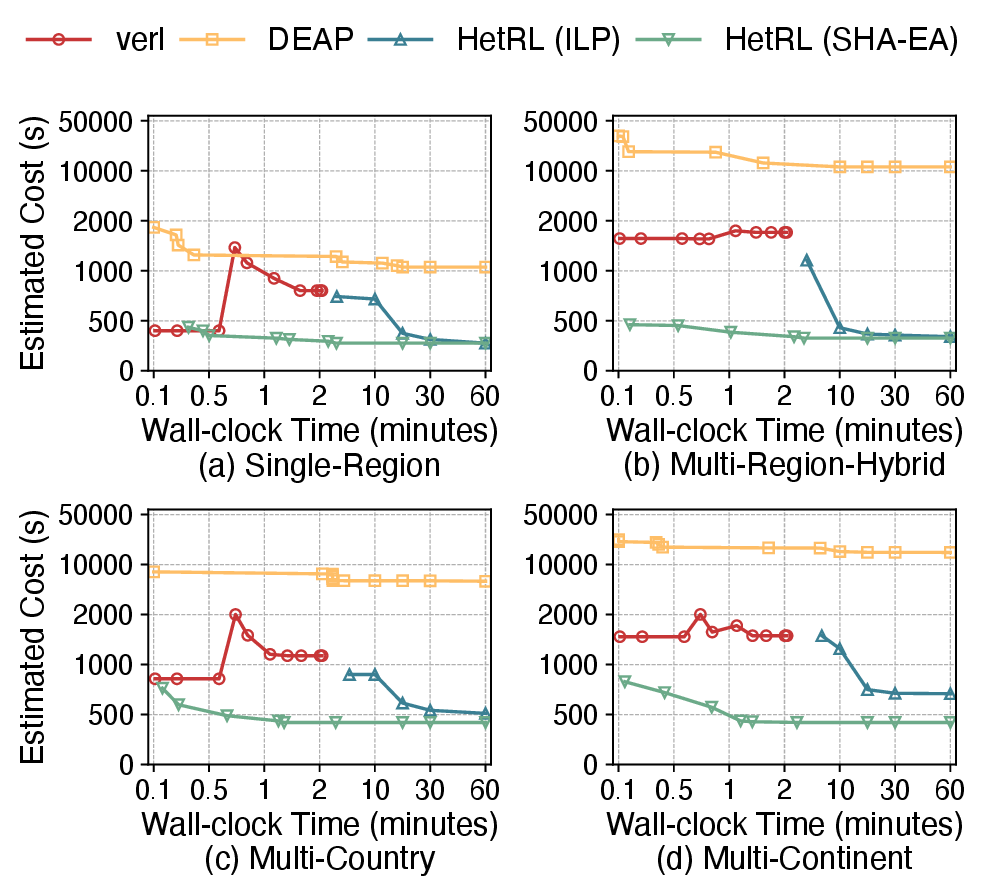}
\caption{Search efficiency comparison across scheduling algorithms for training Qwen-8B with synchronous PPO.}
\label{fig:sched_cost}
\end{figure}

\begin{figure}[t]
\centering
\includegraphics[width=\columnwidth]{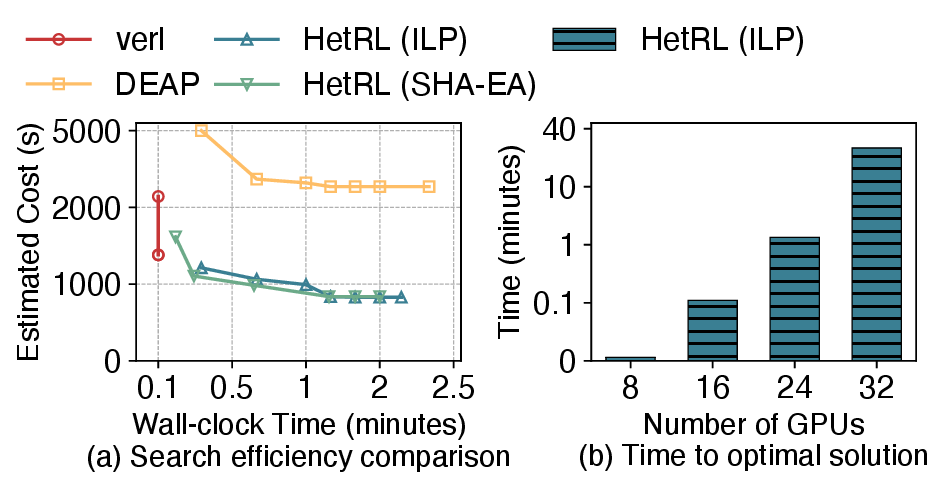}
\caption{Small-scale settings: (a) search efficiency comparison across scheduling algorithms for infrastructures with 24 GPUs; (b) time to optimal solution of \sys(ILP) across different numbers of GPUs.}
\label{fig:sched_cost_small}
\end{figure}

To evaluate the effectiveness of our proposed scheduling algorithms, we compare their search efficiency against verl's scheduling algorithm and a standard evolutionary algorithm implemented using DEAP~\cite{DEAP} for this problem.
The experiments in this section are conducted on a dual-socket server equipped with two 16-core AMD EPYC 9135 CPUs clocked at 3.65GHz, where either multi-threading or multi-processing is employed depending on the experimental variant.
Figure~\ref{fig:sched_cost} shows that our proposed scheduling algorithms significantly outperform verl's scheduling algorithm and DEAP.
When \sys (SHA-EA) and \sys (ILP) are given sufficient search budget (i.e., wall-clock time), both approaches outperform verl due to their heterogeneity-aware cost models, and outperform DEAP due to our proposed design improvements.
However, when the search budget is limited, the plans obtained by \sys (ILP) are worse than those generated by verl, as the ILP-based scheduling algorithm relies on exact solvers that can obtain optimal solutions but incur high computational overhead.
In comparison, \sys (SHA-EA) can efficiently identify near-optimal plans that outperform other variants under different search budgets and achieve comparable quality to those obtained by \sys (ILP) with one hour of search.
For small-scale settings (e.g., no more than 24 GPUs), \sys (ILP) can identify optimal solutions in less than three minutes, as shown in Figure~\ref{fig:sched_cost_small}.
Meanwhile, the performance gaps between the solutions obtained by \sys (SHA-EA) and the optimal solutions obtained by \sys (ILP) are within 1\%.
These results highlight the complementary strengths of the two approaches, enabling practical deployment across diverse settings by leveraging \sys (SHA-EA) for efficient search of near-optimal solutions at scale and \sys (ILP) for exact optimal solutions.

\begin{figure}[t]
\centering
\includegraphics[width=\columnwidth]{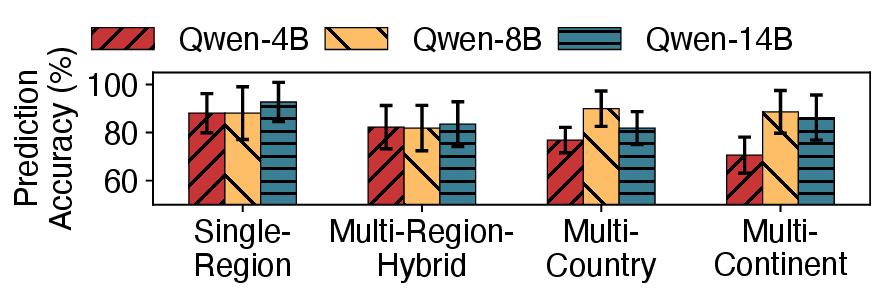}
\caption{Prediction accuracy of the \sys cost model for iteration time across different Qwen model sizes under multiple scenarios. Error bars denote standard deviation.}
\label{fig:cost_model_validation}
\end{figure}

\subsection{Validation of the Cost Model}
We now turn to the validation of our analytical cost model, illustrated in Figure~\ref{fig:cost_model_validation}.
In the Single-Region scenario, the iteration time prediction error of our cost model for RL workloads is comparable to that of existing estimators for pre-training workloads. 
However, in Multi-Region, Multi-Country, and Multi-Continent scenarios, the prediction error is slightly higher due to the increased complexity of RL workloads. 
In addition, the current implementation of RL weight updates in \sys, directly derived from verl, is not highly optimized and may be 100$\times$ slower~\cite{TransferEngine} than the performance upper bound estimated by analytical cost models, further affecting prediction accuracy.
Recent SoTA implementations~\cite{TransferEngine} reduce this gap by leveraging point-to-point communication.

\subsection{Impact of GPU Heterogeneity}
\begin{figure}[t]
\centering
\includegraphics[width=\columnwidth]{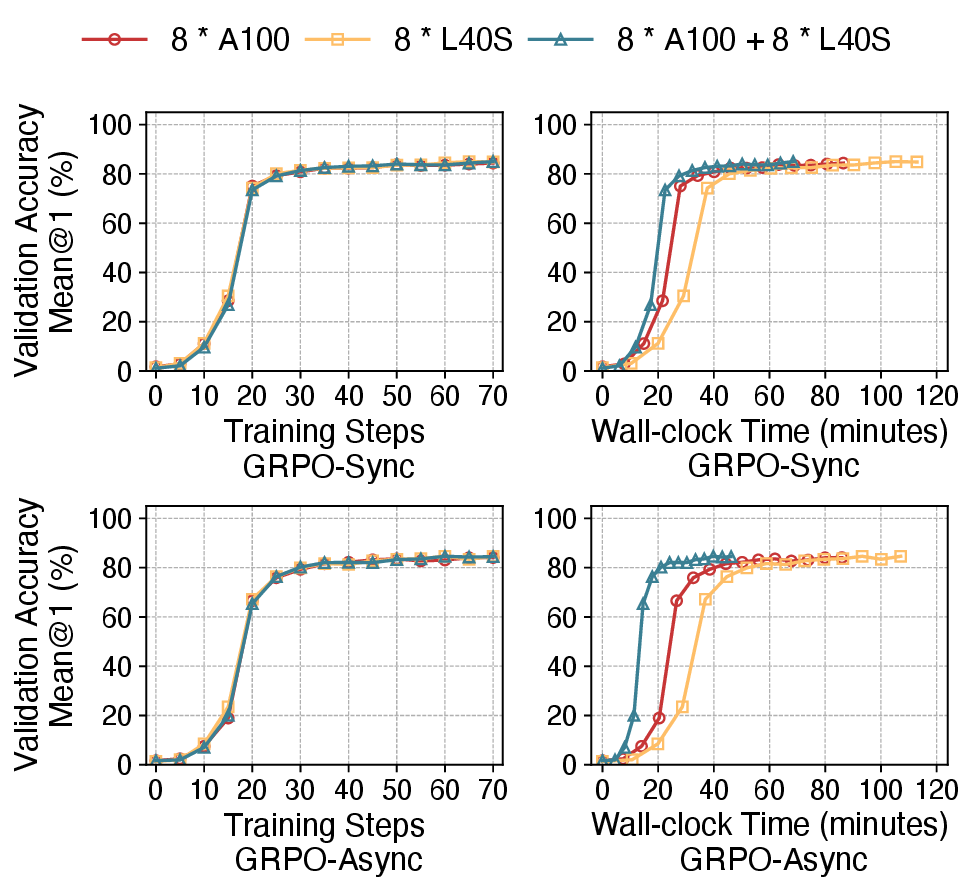}
\caption{Training dynamics of Qwen3-1.7B-Base on GSM8K with GRPO.}
\label{fig:convergence_gsm8k}
\end{figure}

\begin{figure}[t]
\centering
\includegraphics[width=\columnwidth]{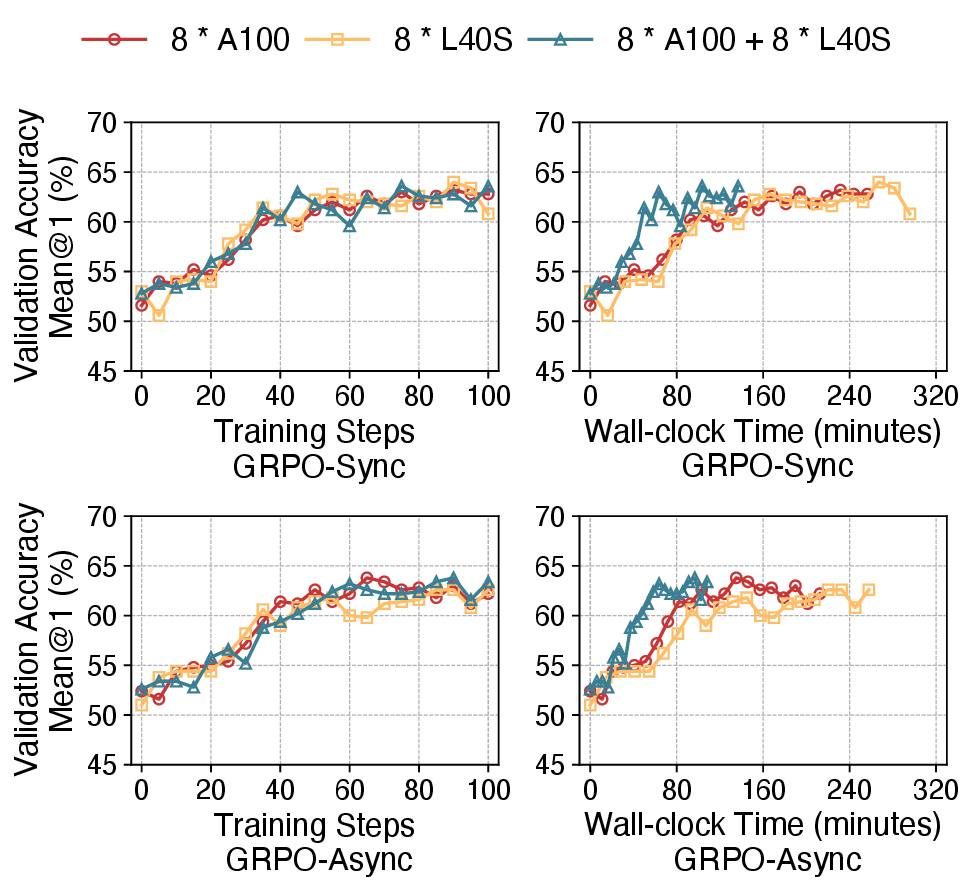}
\caption{Training dynamics of Qwen3-1.7B-Base on MATH-500 with GRPO.}
\label{fig:convergence_math500}
\end{figure}

\textbf{Training quality.}
Now we examine examine the impact of GPU heterogeneity on training quality.
We additionally use Qwen3-1.7B-Base and the MATH-500 dataset, and the rest experimental setup remains consistent with those described in Section~\ref{subsec:setup}.
Although potential precision issues may arise when exchanging data across heterogeneous GPUs, we observe that their impact is negligible for both GRPO-Sync and GRPO-Async across different datasets, as shown in Figures~\ref{fig:convergence_gsm8k} and~\ref{fig:convergence_math500}.
For GRPO-Sync, data exchange across heterogeneous GPUs occurs primarily within each stage, whereas for GRPO-Async, it primarily occurs between stages (i.e., actor training and actor rollout).
When comparing by training steps, using heterogeneous GPUs achieves comparable final validation accuracy.
When comparing by wall-clock time, aggregating the computational and memory capacities of heterogeneous GPUs enables significantly faster convergence.

\begin{figure}[t]
\centering
\includegraphics[width=\columnwidth]{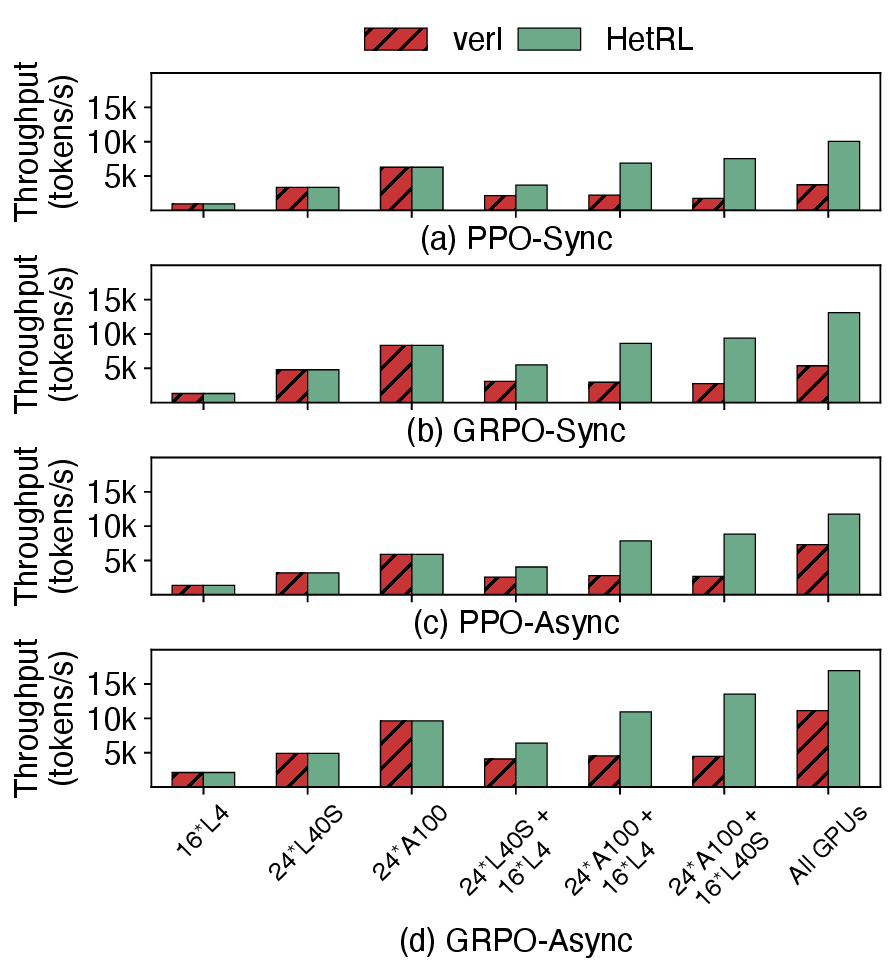}
\caption{Throughput comparison of \sys and verl for training Qwen-8B using different RL algorithms under varying combinations of GPUs.}
\label{fig:gpu_heterogeneity}
\end{figure}

\textbf{Training throughput.}
Our final set of experiments evaluates how different systems perform under varying combinations of heterogeneous GPUs.
\sys outperforms verl by $1.72{\sim}4.33\times$/$1.77{\sim}3.42\times$/$1.57{\sim}3.33\times$/$1.57{\sim}3.03\times$ under PPO-Sync/GRPO-Sync/PPO-Async/GRPO-Async, as shown in Figure~\ref{fig:gpu_heterogeneity}.
When leveraging the larger memory capacity and more compute resources of heterogeneous GPUs (Figure~\ref{fig:gpu_heterogeneity}(ALL GPUs)), \sys outperforms its deployments on limited homogeneous GPUs (Figure~\ref{fig:gpu_heterogeneity}(24*A100)) by $1.57{\sim}2.0\times$.
Due to resource constraints, this set of experiments is limited to the Single-Region scenario.
However, the analysis results obtained from our cost model suggest that the observed improvements can be extended to the rest three scenarios.

Figure~\ref{fig:e2e}(bottom three rows) and Figure~\ref{fig:gpu_heterogeneity}(24$\times$A100) show that, by leveraging heterogeneous compute resources across regions, \sys achieves $1.09{\sim}1.77\times$ the throughput of using only the limited homogeneous GPUs in a single region.
These results show that (1) \sys effectively harnesses heterogeneous compute resources to accelerate RL training; and (2) when available homogeneous GPUs are limited, using heterogeneous resources from the same region or even across regions is a viable alternative.

%% file: 6-discussion.tex
\section{Limitations and Discussion}
\label{sec:discussion}

\textbf{GPU heterogeneity.}
Our evaluation only uses three different NVIDIA GPUs (Table~\ref{tab:gpu}) and relies on AWS OFI NCCL~\cite{aws-ofi-nccl} and AWS EFA~\cite{aws-efa} for inter-node communication.
\sys has not yet been tested on other NVIDIA GPU generations, GPUs from other vendors, or other networking stacks.
We adopt mainstream RL algorithms without modification and focus our evaluation on training throughput.
For training quality, we use a representative RL algorithm (GRPO), an open-source model (Qwen3-1.7B-Base), and datasets (GSM8K and MATH-500) for case studies.
A comprehensive investigation of the impact of potential precision issues arising from data exchange across heterogeneous GPUs on convergence is left for future work.

% \textbf{Prefill-Decode (P/D) disaggregation.}
% Incorporating P/D disaggregation~\cite{DistServe, ThunderServe} into \sys could expand the search space and further improve the performance.
% Our current implementation and the evaluation in this paper do not yet include this feature; however, it can be realized by enabling P/D disaggregation in vLLM~\cite{vLLM} and tweaking our algorithm.

\textbf{Cost-efficiency.}
This work studies accelerating RL training with heterogeneous GPUs and network.
GPU resource pricing varies by vendor, region, and purchasing plan, and is subject to adjustment and fluctuation over time.
Therefore, this paper does not include a cost-efficiency comparison.

\textbf{Simplicity of the scheduling heuristic.}
Compared to non-hybrid alternatives, the proposed hybrid scheduling algorithm introduces additional complexity into \sys.
Below, we discuss why these simpler alternatives cannot achieve comparable performance as shown in Section~\ref{subsec:search_exp}, their connections to our proposed hybrid scheduling algorithm, and additional design considerations that complement Section~\ref{sec:scheduling}.
\begin{itemize}[leftmargin=*,itemsep=-2pt,topsep=-2pt]
    \item Pure SHA. Pure SHA is insufficient for our setting, as it operates only on predefined arms (high-level decisions) and cannot effectively construct overall plans in a large combinatorial space.
    We therefore employ an EA to generate high-quality low-level plans under each high-level decision rather than naive enumeration or sampling.
    \item Pure EA (DEAP). SHA-EA can be viewed as an EA augmented with hierarchical statistical information from SHA, enabling effective early termination of candidates under unpromising high-level decisions.
    In contrast, pure EA relies only on information from the current population, without leveraging broader statistical signals.
    \item Pure ILP. An ILP formulation solved with standard solvers is inefficient in our setting, as our analytical cost model involves deeply nested min–max and max–max operations.
    \item SHA-ILP. While SHA can effectively prune unpromising high-level decisions, using standard solvers for low-level plan construction is impractical, as each arm requires GBs of DRAM and minutes to initialize variables and enforce constraints before solving the problem, especially when the number of GPUs is moderate to large.
    \item Other combinations. SHA and EA are among the simplest algorithms for search budget allocation and plan generation. Using other combinations in the same way does not reduce design or implementation complexity.
\end{itemize}

\textbf{Online redeployment.}
The current design and implementation of \sys do not support online redeployment with fine-grained adaptive mechanisms in highly dynamic environments.
Nevertheless, \sys can accommodate medium-granularity network variability by performing scheduling before the end of the current iteration and during model checkpointing, leveraging the checkpointing capabilities provided by verl.
The updated scheduling plan can then be applied immediately after checkpointing.

\textbf{Multi-tenancy.}
\sys focuses on optimizing the deployment of a single RL training job.
Efficiently scheduling multiple concurrent RL workloads, however, requires different scheduling algorithms and corresponding system designs, some of which are mentioned in Section~\ref{sec:related-work}.

%% file: 7-related-work.tex
\section{Related Work}
\label{sec:related-work}
\textbf{Systems for RL training.}
A series of recent work has focused on optimizing RL training.
~\cite{OpenRLHF, DeepSpeed-Chat, NeMo-Aligner, verl} are pioneering and popular RL training systems that focused on the design of flexible and efficient programming models to support diverse RL workflows when they were proposed, but have continuously adopted system optimizations proposed by subsequent research.
RLHFuse~\cite{RLHFuse} introduces stage fusion strategies to improve GPU utilization for synchronous RL training.
~\cite{AsynchronousRLHF} explores trade-offs between training speed and quality in one-step off-policy, asynchronous RL training and AReaL~\cite{AReaL} proposes staleness-aware asynchronous RL training.
~\cite{LlamaRL, AsyncFlow, StreamRL} provides optimized system support for asynchronous RL training by minimizing bubbles caused by task dependencies.
StreamRL~\cite{StreamRL} confirms the benefits of RL training on cross-data heterogeneous GPUs, but has strict limitations on the heterogeneity of computing resources.
Distinct from existing systems, \sys is the first system tailored for RL training in heterogeneous environments.

\textbf{Heterogeneity-aware LLM training and serving.}
There are plenty of recent efforts have investigated on the deployment of LLM serving and training on heterogeneous GPUs and networks.
ThunderServe~\cite{ThunderServe} formulates heterogeneity-aware LLM serving as job shop scheduling problem and adopts tabu search to identify highly optimized deployment plan.
HexGen~\cite{HexGen} and Helix~\cite{Helix} formulates the problem as a maxflow problem and adopts evolutionary algorithm and mixed integer linear programming algorithm, respectively.
For heterogeneity-aware LLM training, DTFM~\cite{DTFM} formulates it as graph partition problem and also adopt evolutionary algorithm.
Metis~\cite{Metis} considers the deployment on heterogeneous GPUs and homogeneous networks and also adopts depth-first search.
Sailor~\cite{Sailor} and ThunderServe~\cite{ThunderServe} further study fault tolerance and elasticity issues in geo-distributed LLM training and serving and propose heuristic lightweight re-scheduling algorithms.
These systems are targeted for single-model or single-task deployments, but their scheduling optimizations can be integrated as subcomponents into our scheduling algorithm based on the multi-level search framework and successive halving.
Our work shares a similar vision in improving the utilization of geo-distributed heterogeneous GPU resources, but this is the first attempt to tailor a system for RL workflows involving multiple models and tasks with complex computational and data dependencies.

\textbf{Resource scheduling for DL jobs in multi-tenant clusters.}
There are a lot of research on the scheduling optimization for multiple models and multiple tasks.
Gandiva~\cite{Gandiva} explores opportunities for co-locating jobs on the same GPU(s) via temporal sharing, where AntMan~\cite{AntMan} and GSLICE~\cite{GSLICE} colocates traditional DL jobs via spatial sharing.
Extending \sys with optimizations for multi-tenant cluster scheduling is a promising direction for future work.

%% file: 8-conclusion.tex
\section{Conclusion}
\label{sec:conclusion}
In this paper, we explore the opportunity to deploy RL training in infrastructures with heterogeneous GPU models and network characteristics.
Toward this end, we present \sys, a distributed system tailored for such deployments, equipped with (1) a hybrid scheduling algorithm that partitions RL workflows into tasklets, assigns them to heterogeneous computing resources, and efficiently identifies near-optimal solutions, and (2) an ILP-based scheduling algorithm for obtaining optimal solutions.
Our empirical studies suggest that \sys achieves up to 9.17$\times$ higher throughput than SoTA RL training systems under various workloads and settings, demonstrating the effectiveness of our scheduling algorithms and system optimizations.

%% file: appendix.tex
%%%%%%%%%%%%%%%%%%%%%%%%%%%%%%%%%%%%%%%%%%%%%%%%%%%%%%%%%%%%%%%%%%%%%%%%%%%%%%%
%%%%%%%%%%%%%%%%%%%%%%%%%%%%%%%%%%%%%%%%%%%%%%%%%%%%%%%%%%%%%%%%%%%%%%%%%%%%%%%
% SUPPLEMENTAL CONTENT AS APPENDIX AFTER REFERENCES
%%%%%%%%%%%%%%%%%%%%%%%%%%%%%%%%%%%%%%%%%%%%%%%%%%%%%%%%%%%%%%%%%%%%%%%%%%%%%%%
%%%%%%%%%%%%%%%%%%%%%%%%%%%%%%%%%%%%%%%%%%%%%%%%%%%%%%%%%%%%%%%%%%%%%%%%%%%%%%%
\appendix
% \section{Please add supplemental material as appendix here}
% Put anything that you might normally include after the references as an appendix here, {\it not in a separate supplementary file}. Upload your final camera-ready as a single pdf, including all appendices.

%%%%%%%%%%%%%%%%%%%%%%%%%%%%%%%%%%%%%%%%%%%%%%%%%%%%%%%%%%%%%%%%%%%%%%%%%%%%%%%
%%%%%%%%%%%%%%%%%%%%%%%%%%%%%%%%%%%%%%%%%%%%%%%%%%%%%%%%%%%%%%%%%%%%%%%%%%%%%%%
\section{Proof of Proposition~\ref{pro:np_hardness}}
\label{app:proof}
We follow the notation in Section~\ref{subsec:problem}.

\begin{proof}
We prove NP-hardness for the abstract formulation in Section~\ref{subsec:problem} via a theoretical reduction from the \emph{balanced graph partitioning} problem, by constructing a specific instantiation of the objective function $C$.
This construction of the objective function is used solely for complexity analysis.
In contrast, the cost model introduced in Appendix~\ref{app:cost_model} is designed to capture execution behavior in practical systems.

Given an instance of balanced graph partitioning on a weighted undirected graph $H = (U, F)$ where $|U| = k \cdot s$, the goal is to partition $U$ into $k$ disjoint subsets of equal size $s$ while minimizing the cut weight. 
The balanced graph partitioning problem is NP-hard.

We construct a restricted instance of the abstract formulation in Section~\ref{subsec:problem}.
Consider a single RL task with a fixed partitioning strategy $(TP, DP, PP) = (t, d, p)$ such that $t \cdot d \cdot p = |U|$.
This yields $|V_L| = |U|$ tasklets, leaving only the assignment $\sigma: V_L \to V_D$  as the decision variable.

We create $|V_D| = |U|$ devices and partition them into $d$ disjoint groups (islands),  each containing $t \cdot p$ devices.
Let $k := d$ and $s := t \cdot p$, so that $k \cdot s = |U|$.
We set $M_{\mathrm{working}}(l) = 0$, $M_{\mathrm{model}}(l) = 1$, and $M_{\mathrm{gpu}}(d) = 1$ for all tasklets and devices.
Then, by constraints (C2) and (C3), each device hosts exactly one tasklet.

We establish a bijection between vertices $u \in V_D$ and tasklets $l_u \in V_L$.
For each edge $(u,v) \in F$ with weight $w_{uv}$, we define the communication cost as  $0$ if $l_u$ and $l_v$ are assigned to the same island, and $w_{uv}$ otherwise.
The objective of the constructed instance is defined as the total communication cost over all edges in $F$.

Under this construction, any assignment $\sigma$ induces a balanced partition of $U$ into $k$ subsets of size $s$, and its objective value equals the cut weight of the partition.
Conversely, any balanced partition yields a feasible assignment with the same objective value. 

Therefore, the reduction is polynomial-time, and the constructed objective matches the cut weight of the induced partition.
Hence, the restricted scheduling problem is NP-hard.
Since this restricted problem is a special case of the abstract scheduling problem in Section~\ref{subsec:problem}, the general problem is also NP-hard.
\end{proof}

\section{Cost Modeling}
\label{app:cost_model}
In this section, we first model computational and communication costs of the main components involved in training and serving transformer models.
Then we model the cost of the main tasks involved in RL training.
Finally, we model the end-to-end cost for different RL training algorithms.

Since the modeling of memory cost is independent of device and network attributes, we simply follow prior work~\cite{verl,Alpa} and ensure that the tasklets’ memory footprint fits on every assigned device.

\subsection{Notation}
We extend the notation in Section~\ref{sec:scheduling} for device topology graph.
Let $\mathbf{G_D} = (V_D, E_D, \mathbf{comp}, \mathbf{mem}, \mathbf{hbm}, \mathbf{A}, \mathbf{B})$ denote the device topology graph of heterogeneous environments, where $V_D = \{d_1,\ldots,d_N\}$ are a set of $N$ devices and $E_D = V_D \times V_D$ are communication channels between devices.
For each device $d$, the computation capability, memory capacity, and HBM bandwidth are $\mathrm{comp}_d,\mathrm{mem}_d,\mathrm{hbm}_d \in \mathbb{R}_+$.
Collect them into vectors $\mathbf{comp},\mathbf{mem},\mathbf{hbm} \in \mathbb{R}_+^{N}$.
Each edge $e_{d,d^\prime}$ is labeled with latency $\alpha_{d,d^\prime}$ and bandwidth $\beta_{d,d^\prime}$, where $\alpha_{d,d^\prime},\beta_{d,d^\prime} \in \mathbb{R}_+$.
Let $\mathbf{A},\mathbf{B} \in \mathbb{R}_+^{N\times N}$ be the corresponding matrices.

We extend the notation in Section~\ref{sec:scheduling} for tasklet graph.
Let $\mathbf{G_L}^t_{i,j} = ({V_L}^t_{i,j}, {E_L}^t_{i,j})$ denote the subgraph constructed by the set of tasklets of the $j$-th pipeline stage of the $i$-th replica in the $t$-th task.
Let $\mathbf{G_D}^t_{i,j} = ({V_D}^t_{i,j}, {E_D}^t_{i,j})$ be the subgraph constructed by the set of devices assigned to these tasklets and we have $\sigma({V_L}^t_{i,j})={V_D}^t_{i,j}$.
Similarly, let $\mathbf{G_L}^t_{j,k} = ({V_L}^t_{j,k}, {E_L}^t_{j,k})$ denote the subgraph constructed by the set of tasklets of the $k$-th tensor shards of the $j$-th pipeline stage in the $t$-th task and $\mathbf{G_D}^t_{j,k} = ({V_D}^t_{j,k}, {E_D}^t_{j,k})$ denote the assigned devices.

Now we introduce other required notation.
Let $h_1^t,h_2^t,nl^t$ be the hidden size, the intermediate size, and number of layers of the LLM of the $t$-th task.
Let $nl_j^t, \mathrm{TP}_j^t$ be number of layers and TP degree in $j$-th pipeline stage of the $t$-th task.
Let $\mathrm{seq}_\mathrm{in}$ and $\mathrm{seq}_\mathrm{out}$ denote the maximum sequence lengths of the input prompts and output responses.
Let $mbs,nm$ denote micro-batch size and number of micro-batches.
Please note that we have preprocessed $nm$ based on the number of responses generated per prompt, the data parallelism degree.
Let $B_{\mathrm{BF16}}, B_{\mathrm{FP32}}$ be the data sizes of the BF16 and FP32 data types.
Let $\mathrm{ring}$ denote a function that returns the set of all feasible ring graphs of given devices.
For brevity, we have omitted the vocabulary and token embeddings in the cost model, but they are included in our actual implementation.

\subsection{Modeling component-level computation and communication costs.}
\textbf{Tensor parallelism communication.}
The communication volume transferred between a pair of neighboring GPUs in $\mathbf{G_D}^t_{i,j}$ per all-reduce during TP can be estimated by:
\begin{align*}
    \mathrm{cv}_\mathrm{tp}(t,i,j) &= B_\mathrm{BF16} \cdot mbs \cdot (\mathrm{seq}_\mathrm{in}+\mathrm{seq}_\mathrm{out}) \cdot h_1^t \\
    &\quad \cdot \frac{2 \cdot (\mathrm{TP}_j^t - 1)}{\mathrm{TP}_j^t}
\end{align*}
The TP communication cost of forward passes for $\mathbf{G_L}^t_{i,j}$ can be estimated by:
\begin{align*}
    C_\mathrm{tp}(t,i,j) &= 2 \cdot nm \cdot nl_j^t \\
    &\quad \cdot \min_{r \in \mathrm{ring}(\mathbf{G_D}^t_{i,j})} \max_{e_{d,d^\prime} \in r} (\alpha_{d,d^\prime}+\frac{\mathrm{cv}_\mathrm{tp}(t,i,j)}{\beta_{d,d^\prime}})
\end{align*}
If recomputation is enabled, the TP communication cost of both forward and backward passes for $\mathbf{G_L}^t_{i,j}$ can be estimated by:
\begin{align*}
    C_\mathrm{tp}(t,i,j) &= 6 \cdot nm \cdot nl_j^t \\
    &\quad \cdot \min_{r \in \mathrm{ring}(\mathbf{G_D}^t_{i,j})} \max_{e_{d,d^\prime} \in r} (\alpha_{d,d^\prime}+\frac{\mathrm{cv}_{tp}(t,i,j)}{\beta_{d,d^\prime}})
\end{align*}
The overall TP communication cost can be estimated by:
\begin{align*}
    C_\mathrm{tp}^t = \max_{i,j} C_\mathrm{tp}(t,i,j)
\end{align*}

\textbf{Pipeline parallelism communication.}
The communication volume transferred between the $j$-th and $j+1$-th pipeline stages per micro-batch during PP can be estimated by:
\begin{align*}
    \mathrm{cv}_\mathrm{pp}(t,i,j) = B_\mathrm{BF16} \cdot mbs \cdot (\mathrm{seq}_\mathrm{in}+\mathrm{seq}_\mathrm{out}) \cdot h_1^t
\end{align*}
The PP communication cost of forward passes between $\mathbf{G_L}^t_{i,j}$ and $\mathbf{G_L}^t_{i,j+1}$ can be   by:
\begin{align*}
    C_\mathrm{pp}(t,i,j) &= nm \\
    &\quad \cdot \min_{d \in {V_D}^t_{i,j}, d^\prime \in {V_D}^t_{i,j+1}} (\alpha_{d,d^\prime}+\frac{\mathrm{cv}_\mathrm{pp}(t,i,j)}{\beta_{d,d^\prime}})
\end{align*}
The PP communication cost of both forward and backward passes between $\mathbf{G_L}^t_{i,j}$ and $\mathbf{G_L}^t_{i,j+1}$ can be estimated by:
\begin{align*}
    C_\mathrm{pp}(t,i,j) &= 2 \cdot nm \\
    &\quad \cdot \min_{d \in {V_D}^t_{i,j}, d^\prime \in {V_D}^t_{i,j+1}} (\alpha_{d,d^\prime}+\frac{\mathrm{cv}_\mathrm{pp}(t,i,j)}{\beta_{d,d^\prime}})
\end{align*}
The overall PP communication cost can be estimated by:
\begin{align*}
    C_\mathrm{pp}^t = \max_{i,j} C_\mathrm{pp}(t,i,j)
\end{align*}

\textbf{Data parallelism communication.}
The communication volume transferred between a pair of neighboring GPUs in $\mathbf{G_D}^t_{j,k}$ per all-reduce during DP can be estimated by:
\begin{align*}
    \mathrm{cv}_\mathrm{dp}(t,j,k) &= B_\mathrm{BF16} \cdot nl_j^t \cdot (4 \cdot {h_1^t}^2 + 3 \cdot h_1^t \cdot h_2^t) \\
    &\quad \cdot \frac{2 \cdot (|{V_L}^t_{j,k}| - 1)}{|{V_L}^t_{j,k}|\cdot\mathrm{TP}_j^t}
\end{align*}
The DP communication cost for $\mathbf{G_D}^t_{j,k}$ per iteration can be estimated by:
\begin{align*}
    C_\mathrm{dp}(t,j,k) = \min_{r \in \mathrm{ring}(\mathbf{G_D}^t_{j,k})} \max_{e_{d,d^\prime} \in r} (\alpha_{d,d^\prime}+\frac{\mathrm{cv}_\mathrm{dp}(t,j,k)}{\beta_{d,d^\prime}})
\end{align*}
The overall DP communication cost can be estimated by:
\begin{align*}
    C_\mathrm{dp}^t = \max_{j,k} C_\mathrm{dp}(t,j,k)
\end{align*}

\textbf{Computation.}
The computational cost of a transformer layer per forward pass per sample can be estimated by:
\begin{align*}
    & C_\mathrm{comp}^{\mathrm{qkvo\_proj}}(t,i,j,k) = 2 \cdot 4 \cdot (\mathrm{seq}_\mathrm{in}+\mathrm{seq}_\mathrm{out}) \cdot {h_1^t}^2 \\
    & C_\mathrm{comp}^{\mathrm{attn}}(t,i,j,k) = 2 \cdot 2 \cdot (\mathrm{seq}_\mathrm{in}+\mathrm{seq}_\mathrm{out})^2 \cdot h_1^t \\
    & C_\mathrm{comp}^{\mathrm{mlp}}(t,i,j,k) = 2 \cdot 3 \cdot (\mathrm{seq}_\mathrm{in}+\mathrm{seq}_\mathrm{out}) \cdot h_1^t \cdot h_2^t \\
    & C_\mathrm{comp}^{\mathrm{layer}}(t,i,j,k) = C_\mathrm{comp}^{\mathrm{qkvo\_proj}} + C_\mathrm{comp}^{\mathrm{attn}} + C_\mathrm{comp}^{\mathrm{mlp}}
\end{align*}
The computation cost of forward passes for $l^t_{i,j,k}$ on device $d=\sigma(l^t_{i,j,k})$ can be estimated by:
\begin{align*}
    C_\mathrm{comp}(t,i,j,k) = nm \cdot mbs \cdot nl_j^t \cdot \frac{C_\mathrm{comp}^{\mathrm{layer}}}{\mathrm{comp}_d \cdot \mathrm{TP}_j^t}
\end{align*}
The computation cost of both forward and backward passes for $l^t_{i,j,k}$ on device $d=\sigma(l^t_{i,j,k})$ can be estimated by:
\begin{align*}
    C_\mathrm{comp}(t,i,j,k) = 3 \cdot nm \cdot mbs \cdot nl_j^t \cdot \frac{C_\mathrm{comp}^{\mathrm{layer}}}{\mathrm{comp}_d \cdot \mathrm{TP}_j^t}
\end{align*}
The computation cost for $\mathbf{G_L}^t_{i,j}$ can be estimated by:
\begin{align*}
    C_\mathrm{comp}(t,i,j) = \max_{k} C_\mathrm{comp}(t,i,j,k)
\end{align*}
The overall computation cost can be estimated by:
\begin{align*}
    C_\mathrm{comp}^t = \max_{i,j} C_\mathrm{comp}(t,i,j)
\end{align*}

Please note that here we set $\mathrm{seq}_\mathrm{out}$ to $0$ in the estimation for the actor generation task.

\textbf{Pipeline bubbles.}
The cost of pipeline bubbles for the $i$-th replica can be estimated by:
\begin{align*}
    C_\mathrm{bubble}(t, i) &= \sum_{j \neq 0} \\
    &\quad \frac{C_\mathrm{comp}(t,i,j) + C_\mathrm{tp}(t,i,j) + C_\mathrm{pp}(t,i,j)}{nm}
\end{align*}
The overall cost of pipeline bubbles can be estimated by:
\begin{align*}
    C_\mathrm{bubble}^t = \max_i C_\mathrm{bubble}(t, i)
\end{align*}

\textbf{Decoding (HBM-bandwidth bound).}
The hbm cost for $l^t_{i,j,k}$ on device $d=\sigma(l^t_{i,j,k})$ can be estimated by:
\begin{align*}
    C_\mathrm{hbm}(t,i,j,k) &= \mathrm{seq}_\mathrm{out} \cdot nm \cdot mbs \\
    &\quad \cdot \frac{B_\mathrm{BF16} \cdot nl_j^t \cdot (4 \cdot {h_1^t}^2 + 3 \cdot h_1^t \cdot h_2^t)}{dbs_d \cdot \mathrm{hbm}_d\cdot\mathrm{TP}_j^t},
\end{align*}
where $dbs_d$ is the decoding batch size of the LLM serving engine deployed on device $d$.
The hbm cost for the $i$-th replica can be estimated by:
\begin{align*}
    C_\mathrm{hbm}(t, i) = \max_{j,k} C_\mathrm{hbm}(t,i,j,k)
\end{align*}
The overall hbm cost can be estimated by:
\begin{align*}
    C_\mathrm{hbm}^t = \max_i C_\mathrm{hbm}(t,i)
\end{align*}

\textbf{Resharding.}
The communication volume for all-gather during model resharding between a pair of neighboring GPUs in the $i$-th model replica group of the actor training task $t$ can be estimated by:
\begin{align*}
    \mathrm{cv}_\mathrm{all-gather}(t,i) &= B_\mathrm{BF16} \cdot nl^t \cdot (4 \cdot {h_1^t}^2 + 3 \cdot h_1^t \cdot h_2^t) \\
    &\quad \cdot \frac{|\mathbf{V_L}^t_i| - 1}{|\mathbf{V_L}^t_i|}
\end{align*}

The all-gather cost for $\mathbf{G_D}^t_i$ per iteration can be estimated by:
\begin{align*}
    C_\mathrm{all-gather}(t,i) =& \min_{r \in \mathrm{ring}(\mathbf{V_D}^t_i)} \\
    & \max_{e_{d,d^\prime} \in r} (\alpha_{d,d^\prime}+\frac{\mathrm{cv}_\mathrm{all-gather}(t,i)}{\beta_{d,d^\prime}})
\end{align*}
The naive overall model resharding cost required for synchronous RL training can be estimated by:
\begin{align*}
    C_\mathrm{reshard} = \max_{i} C_\mathrm{all-gather}(t,i)
\end{align*}

\textbf{Synchronization.}
The communication volume for all-gather during weight synchronization between a pair of neighboring GPUs in the $i$-th model replica group of actor training task $t$ can be estimated by:
\begin{align*}
    \mathrm{cv}_\mathrm{all-gather}(t,i) &= B_\mathrm{BF16} \cdot nl^t \cdot (4 \cdot {h_1^t}^2 + 3 \cdot h_1^t \cdot h_2^t) \\
    &\quad \cdot \frac{|\mathbf{V_L}^t_i| - 1}{|\mathbf{V_L}^t_i|}
\end{align*}

The all-gather cost for $\mathbf{G_D}^t_i$ per iteration can be estimated by:
\begin{align*}
    C_\mathrm{all-gather}(t,i) =& \min_{r \in \mathrm{ring}(\mathbf{V_D}^t_i)} \\
    & \max_{e_{d,d^\prime} \in r} (\alpha_{d,d^\prime}+\frac{\mathrm{cv}_\mathrm{all-gather}(t,i)}{\beta_{d,d^\prime}})
\end{align*}

The communication volume for broadcast during weight synchronization between a pair of neighboring GPUs in the $i^\prime$-th model replica group of actor generation task $t^\prime$ can be estimated by:
\begin{align*}
    \mathrm{cv}_\mathrm{broadcast}(t^\prime,i^\prime) &= B_\mathrm{BF16} \cdot nl^{t^\prime} \cdot (4 \cdot {h_1^{t^\prime}}^2 + 3 \cdot h_1^{t^\prime} \cdot h_2^{t^\prime}) \\
    &\quad \cdot \frac{|\mathbf{V_L}^t_i| - 1}{|\mathbf{V_L}^t_i|}
\end{align*}

The broadcast cost for $\mathbf{G_D}^{t^\prime}_{i^\prime}$ per iteration can be estimated by:
\begin{align*}
    C_\mathrm{broadcast}({t^\prime},{i^\prime}) =& \min_{r \in \mathrm{ring}(\mathbf{V_D}^{t^\prime}_{i^\prime})} \\
    & \max_{e_{d,d^\prime} \in r} (\alpha_{d,d^\prime}+\frac{\mathrm{cv}_\mathrm{all-gather}({t^\prime},{i^\prime})}{\beta_{d,d^\prime}})
\end{align*}

The communication volume for naive point-to-point communication during weight synchronization between a GPU assigned to actor training task $t$ and another GPU assigned to actor generation task $t^\prime$ can be estimated by:
\begin{align*}
    \mathrm{cv}_\mathrm{p2p}(t,t^\prime) &= B_\mathrm{BF16} \cdot nl^t \cdot (4 \cdot {h_1^t}^2 + 3 \cdot h_1^t \cdot h_2^t)
\end{align*}

The point-to-point communication cost for $\mathbf{G_D}^{t^\prime}_{i^\prime}$ per iteration can be estimated by:
\begin{align*}
    C_\mathrm{p2p}(t,{t^\prime}) = \min_{d \in |V_D^t|, d^\prime \in |V_D^{t^\prime}|} (\alpha_{d,d^\prime}+\frac{\mathrm{cv}_\mathrm{p2p}({t^\prime},{i^\prime})}{\beta_{d,d^\prime}})
\end{align*}

The naive overall weight synchronization cost required for synchronous RL training can be estimated by:
\begin{align*}
    C_\mathrm{sync} =& \min_{i} C_\mathrm{all-gather}(t,i) +  \max_{i^\prime} C_\mathrm{broadcast}(t^\prime,i^\prime) \\
    &+ C_\mathrm{p2p}(t,{t^\prime})
\end{align*}

\subsection{Modeling task-level costs.}
One can simply add up the overall costs provided by the above equations for estimation.
However, for more fine-grained estimates, the cost models are as follows:

\textbf{Generation.} The cost of generation task can be estimated by:
\begin{align*}
    \Psi^\mathrm{gen}&(C_\mathrm{comp}^t, C_\mathrm{tp}^t, C_\mathrm{pp}^t, C_\mathrm{hbm}^t) \\
    & = \max_{i} (\max_{j} (C_\mathrm{comp}(t,i,j) + C_\mathrm{tp}(t,i,j) \\
    &\quad + C_\mathrm{pp}(t,i,j) + C_\mathrm{hbm}(t, i, j)))
\end{align*}

\textbf{Inference.} The cost of inference tasks can be estimated by:
\begin{align*}
    \Psi^\mathrm{inf} &(C_\mathrm{comp}^t, C_\mathrm{tp}^t, C_\mathrm{pp}^t) \\
    & = \max_{i} (\max_{j} (C_\mathrm{comp}(t,i,j) + C_\mathrm{tp}(t,i,j)\\
        &\quad + C_\mathrm{pp}(t,i,j))) \\
\end{align*}

\textbf{Training.} The cost of training tasks can be estimated by:
\begin{align*}
    \Psi^\mathrm{train} &(C_\mathrm{comp}^t, C_\mathrm{tp}^t, C_\mathrm{pp}^t, C_\mathrm{dp}^t, C_\mathrm{bubble}^t) \\
    & = \max_{i} (\max_{j} (C_\mathrm{comp}(t,i,j) + C_\mathrm{tp}(t,i,j) \\
    &\quad + C_\mathrm{pp}(t,i,j)) + C_\mathrm{bubble}(t, i))+ C_\mathrm{dp}^t \\
\end{align*}

\iffalse
\textbf{Resharding.}
The overall communication cost of naive resharding per iteration can be estimated by:
\begin{align*}
    C_\mathrm{reshard}^t = \\
\end{align*}

\textbf{Weight synchronization.}
The overall communication cost of naive weight synchronization per iteration can be estimated by:
\begin{align*}
    C_\mathrm{sync}^t = \\
\end{align*}

We leave the optimization of sharding and weight synchronization in heterogeneous environments as a future research direction.
\fi
For load balancing strategies, one can tune $\mathrm{seq}_\mathrm{in}, mb, nl_j^t$ during cost modeling.

\subsection{Modeling end-to-end costs.}
Let $C^{1:6}(\cdot)$ denote the cost models of actor generation, reward inference, reference inference, critic inference, actor training, and critic training, respectively, and $\Phi(\cdot)$ denotes the cost of the tasks without dependencies which can be defined as
\begin{align*}
    \Phi(\{C^t\}) &= \max_t C^t + (1 - \eta)\!\left(\sum_t C^t - \max_t C^t\right),\\
    &\quad \eta \in [0,1],
\end{align*}
where the coefficient $\eta$ parameterizes the level of task parallelism (0: sequential; 1: fully parallel; else: partial).

The cost mode of $t$-th task is given by:
\begin{align*}
    C^t =
    \begin{cases}
        \Psi^\mathrm{gen}(C_\mathrm{comp}^t, C_\mathrm{hbm}^t, C_\mathrm{tp}^t, C_\mathrm{pp}^t), & t = 1,\\
        \Psi^\mathrm{inf}(C_\mathrm{comp}^t, C_\mathrm{tp}^t, C_\mathrm{pp}^t), & t \in \{2,3,4\},\\
        \Psi^\mathrm{train}(C_\mathrm{comp}^t, C_\mathrm{tp}^t, C_\mathrm{pp}^t, C_\mathrm{dp}^t, \\
        \qquad C_\mathrm{bubble}^t), & t \in \{5,6\},
    \end{cases}
\end{align*}

Finally, the cost of different RL training algorithms can be estimated as below:

\textbf{Synchronous PPO.}
\begin{align*}
    C_\mathrm{SyncPPO} &= C^1 + \Phi(\{C^2, C^3, C^4\}) + \Phi(\{C^5, C^6\}) \\
    &\quad + C_\mathrm{reshard}
\end{align*}

\textbf{Asynchronous PPO.}
\begin{align*}
    C_\mathrm{AsyncPPO} &= \max (C^1, \Phi(\{C^2, C^3, C^4\}) +\\
    &\quad \Phi(\{C^5, C^6\})) + C_\mathrm{sync}
\end{align*}

\textbf{Synchronous GRPO.}
\begin{align*}
    C_\mathrm{SyncGRPO} = C^1 + \Phi(\{C^2, C^3\}) + C^6 +  + C_\mathrm{reshard}
\end{align*}

\textbf{Asynchronous GRPO.}
\begin{align*}
    C_\mathrm{AsyncGRPO} = \max (C^1, \Phi(\{C^2, C^3\}) + C^6) + C_\mathrm{sync}
\end{align*}

%% file: example_paper.bib
@misc{aws-efa,
    author = {},
    title = {AWS Elastic Fabric Adapter},
    year = {2025},
    organization = {Amazon Web Services},
    howpublished = {\url{https://aws.amazon.com/hpc/efa/}},
}

@misc{aws-ofi-nccl,
    author = {},
    title = {AWS OFI NCCL},
    year = {2025},
    organization = {Amazon Web Services},
    howpublished = {\url{https://github.com/aws/aws-ofi-nccl}},
}

@article{TransferEngine,
  author       = {N{\'{a}}ndor Licker and
                  Kevin Hu and
                  Vladimir Zaytsev and
                  Lequn Chen},
  title        = {{RDMA} Point-to-Point Communication for {LLM} Systems},
  journal      = {CoRR},
  volume       = {abs/2510.27656},
  year         = {2025},
  url          = {https://doi.org/10.48550/arXiv.2510.27656},
  doi          = {10.48550/ARXIV.2510.27656},
  eprinttype   = {arXiv},
  eprint       = {2510.27656},
  bibsource    = {dblp computer science bibliography, https://dblp.org}
}

@inproceedings{verl,
  author       = {Guangming Sheng and
                  Chi Zhang and
                  Zilingfeng Ye and
                  Xibin Wu and
                  Wang Zhang and
                  Ru Zhang and
                  Yanghua Peng and
                  Haibin Lin and
                  Chuan Wu},
  title        = {HybridFlow: {A} Flexible and Efficient {RLHF} Framework},
  booktitle    = {Proceedings of the Twentieth European Conference on Computer Systems,
                  EuroSys 2025, Rotterdam, The Netherlands, 30 March 2025 - 3 April
                  2025},
  pages        = {1279--1297},
  publisher    = {{ACM}},
  year         = {2025},
  url          = {https://doi.org/10.1145/3689031.3696075},
  doi          = {10.1145/3689031.3696075},
  bibsource    = {dblp computer science bibliography, https://dblp.org}
}

@inproceedings{RLHFuse,
  author       = {Yinmin Zhong and
                  Zili Zhang and
                  Bingyang Wu and
                  Shengyu Liu and
                  Yukun Chen and
                  Changyi Wan and
                  Hanpeng Hu and
                  Lei Xia and
                  Ranchen Ming and
                  Yibo Zhu and
                  Xin Jin},
  editor       = {Theophilus A. Benson and
                  Radhika Niranjan Mysore},
  title        = {Optimizing {RLHF} Training for Large Language Models with Stage Fusion},
  booktitle    = {22nd {USENIX} Symposium on Networked Systems Design and Implementation,
                  {NSDI} 2025, Philadelphia, PA, USA, April 28-30, 2025},
  pages        = {489--503},
  publisher    = {{USENIX} Association},
  year         = {2025},
  url          = {https://www.usenix.org/conference/nsdi25/presentation/zhong},
  bibsource    = {dblp computer science bibliography, https://dblp.org}
}

@article{AReaL,
  author       = {Wei Fu and
                  Jiaxuan Gao and
                  Xujie Shen and
                  Chen Zhu and
                  Zhiyu Mei and
                  Chuyi He and
                  Shusheng Xu and
                  Guo Wei and
                  Jun Mei and
                  Jiashu Wang and
                  Tongkai Yang and
                  Binhang Yuan and
                  Yi Wu},
  title        = {AReaL: {A} Large-Scale Asynchronous Reinforcement Learning System
                  for Language Reasoning},
  journal      = {CoRR},
  volume       = {abs/2505.24298},
  year         = {2025},
  url          = {https://doi.org/10.48550/arXiv.2505.24298},
  doi          = {10.48550/ARXIV.2505.24298},
  eprinttype    = {arXiv},
  eprint       = {2505.24298},
  bibsource    = {dblp computer science bibliography, https://dblp.org}
}

@article{StreamRL,
  author       = {Yinmin Zhong and
                  Zili Zhang and
                  Xiaoniu Song and
                  Hanpeng Hu and
                  Chao Jin and
                  Bingyang Wu and
                  Nuo Chen and
                  Yukun Chen and
                  Yu Zhou and
                  Changyi Wan and
                  Hongyu Zhou and
                  Yimin Jiang and
                  Yibo Zhu and
                  Daxin Jiang},
  title        = {StreamRL: Scalable, Heterogeneous, and Elastic {RL} for LLMs with
                  Disaggregated Stream Generation},
  journal      = {CoRR},
  volume       = {abs/2504.15930},
  year         = {2025},
  url          = {https://doi.org/10.48550/arXiv.2504.15930},
  doi          = {10.48550/ARXIV.2504.15930},
  eprinttype    = {arXiv},
  eprint       = {2504.15930},
  bibsource    = {dblp computer science bibliography, https://dblp.org}
}

@article{LlamaRL,
  author       = {Bo Wu and
                  Sid Wang and
                  Yunhao Tang and
                  Jia Ding and
                  Eryk Helenowski and
                  Liang Tan and
                  Tengyu Xu and
                  Tushar Gowda and
                  Zhengxing Chen and
                  Chen Zhu and
                  Xiaocheng Tang and
                  Yundi Qian and
                  Beibei Zhu and
                  Rui Hou},
  title        = {LlamaRL: {A} Distributed Asynchronous Reinforcement Learning Framework
                  for Efficient Large-scale {LLM} Training},
  journal      = {CoRR},
  volume       = {abs/2505.24034},
  year         = {2025},
  url          = {https://doi.org/10.48550/arXiv.2505.24034},
  doi          = {10.48550/ARXIV.2505.24034},
  eprinttype    = {arXiv},
  eprint       = {2505.24034},
  bibsource    = {dblp computer science bibliography, https://dblp.org}
}

@article{AsyncFlow,
  author       = {Zhenyu Han and
                  Ansheng You and
                  Haibo Wang and
                  Kui Luo and
                  Guang Yang and
                  Wenqi Shi and
                  Menglong Chen and
                  Sicheng Zhang and
                  Zeshun Lan and
                  Chunshi Deng and
                  Huazhong Ji and
                  Wenjie Liu and
                  Yu Huang and
                  Yixiang Zhang and
                  Chenyi Pan and
                  Jing Wang and
                  Xin Huang and
                  Chunsheng Li and
                  Jianping Wu},
  title        = {AsyncFlow: An Asynchronous Streaming {RL} Framework for Efficient
                  {LLM} Post-Training},
  journal      = {CoRR},
  volume       = {abs/2507.01663},
  year         = {2025},
  url          = {https://doi.org/10.48550/arXiv.2507.01663},
  doi          = {10.48550/ARXIV.2507.01663},
  eprinttype    = {arXiv},
  eprint       = {2507.01663},
  bibsource    = {dblp computer science bibliography, https://dblp.org}
}

@inproceedings{AsynchronousRLHF,
  author       = {Michael Noukhovitch and
                  Shengyi Huang and
                  Sophie Xhonneux and
                  Arian Hosseini and
                  Rishabh Agarwal and
                  Aaron C. Courville},
  title        = {Asynchronous {RLHF:} Faster and More Efficient Off-Policy {RL} for
                  Language Models},
  booktitle    = {The Thirteenth International Conference on Learning Representations,
                  {ICLR} 2025, Singapore, April 24-28, 2025},
  publisher    = {OpenReview.net},
  year         = {2025},
  url          = {https://openreview.net/forum?id=FhTAG591Ve},
  bibsource    = {dblp computer science bibliography, https://dblp.org}
}

@article{OpenRLHF,
  author       = {Jian Hu and
                  Xibin Wu and
                  Weixun Wang and
                  Xianyu and
                  Dehao Zhang and
                  Yu Cao},
  title        = {OpenRLHF: An Easy-to-use, Scalable and High-performance {RLHF} Framework},
  journal      = {CoRR},
  volume       = {abs/2405.11143},
  year         = {2024},
  url          = {https://doi.org/10.48550/arXiv.2405.11143},
  doi          = {10.48550/ARXIV.2405.11143},
  eprinttype    = {arXiv},
  eprint       = {2405.11143},
  bibsource    = {dblp computer science bibliography, https://dblp.org}
}

@article{NeMo-Aligner,
  author       = {Gerald Shen and
                  Zhilin Wang and
                  Olivier Delalleau and
                  Jiaqi Zeng and
                  Yi Dong and
                  Daniel Egert and
                  Shengyang Sun and
                  Jimmy J. Zhang and
                  Sahil Jain and
                  Ali Taghibakhshi and
                  Markel Sanz Ausin and
                  Ashwath Aithal and
                  Oleksii Kuchaiev},
  title        = {NeMo-Aligner: Scalable Toolkit for Efficient Model Alignment},
  journal      = {CoRR},
  volume       = {abs/2405.01481},
  year         = {2024},
  url          = {https://doi.org/10.48550/arXiv.2405.01481},
  doi          = {10.48550/ARXIV.2405.01481},
  eprinttype    = {arXiv},
  eprint       = {2405.01481},
  bibsource    = {dblp computer science bibliography, https://dblp.org}
}

@article{DeepSpeed-Chat,
  author       = {Zhewei Yao and
                  Reza Yazdani Aminabadi and
                  Olatunji Ruwase and
                  Samyam Rajbhandari and
                  Xiaoxia Wu and
                  Ammar Ahmad Awan and
                  Jeff Rasley and
                  Minjia Zhang and
                  Conglong Li and
                  Connor Holmes and
                  Zhongzhu Zhou and
                  Michael Wyatt and
                  Molly Smith and
                  Lev Kurilenko and
                  Heyang Qin and
                  Masahiro Tanaka and
                  Shuai Che and
                  Shuaiwen Leon Song and
                  Yuxiong He},
  title        = {DeepSpeed-Chat: Easy, Fast and Affordable {RLHF} Training of ChatGPT-like
                  Models at All Scales},
  journal      = {CoRR},
  volume       = {abs/2308.01320},
  year         = {2023},
  url          = {https://doi.org/10.48550/arXiv.2308.01320},
  doi          = {10.48550/ARXIV.2308.01320},
  eprinttype    = {arXiv},
  eprint       = {2308.01320},
  bibsource    = {dblp computer science bibliography, https://dblp.org}
}

@article{DBLP:journals/tmlr/KaufmannWBH25,
  author       = {Timo Kaufmann and
                  Paul Weng and
                  Viktor Bengs and
                  Eyke H{\"{u}}llermeier},
  title        = {A Survey of Reinforcement Learning from Human Feedback},
  journal      = {Trans. Mach. Learn. Res.},
  volume       = {2025},
  year         = {2025},
  url          = {https://openreview.net/forum?id=f7OkIurx4b},
  bibsource    = {dblp computer science bibliography, https://dblp.org}
}

@article{Tulu3,
  author       = {Nathan Lambert and
                  Jacob Morrison and
                  Valentina Pyatkin and
                  Shengyi Huang and
                  Hamish Ivison and
                  Faeze Brahman and
                  Lester James V. Miranda and
                  Alisa Liu and
                  Nouha Dziri and
                  Shane Lyu and
                  Yuling Gu and
                  Saumya Malik and
                  Victoria Graf and
                  Jena D. Hwang and
                  Jiangjiang Yang and
                  Ronan Le Bras and
                  Oyvind Tafjord and
                  Chris Wilhelm and
                  Luca Soldaini and
                  Noah A. Smith and
                  Yizhong Wang and
                  Pradeep Dasigi and
                  Hannaneh Hajishirzi},
  title        = {T{\"{U}}LU 3: Pushing Frontiers in Open Language Model Post-Training},
  journal      = {CoRR},
  volume       = {abs/2411.15124},
  year         = {2024},
  url          = {https://doi.org/10.48550/arXiv.2411.15124},
  doi          = {10.48550/ARXIV.2411.15124},
  eprinttype    = {arXiv},
  eprint       = {2411.15124},
  bibsource    = {dblp computer science bibliography, https://dblp.org}
}

@article{GRPO,
  author       = {Zhihong Shao and
                  Peiyi Wang and
                  Qihao Zhu and
                  Runxin Xu and
                  Junxiao Song and
                  Mingchuan Zhang and
                  Y. K. Li and
                  Y. Wu and
                  Daya Guo},
  title        = {DeepSeekMath: Pushing the Limits of Mathematical Reasoning in Open
                  Language Models},
  journal      = {CoRR},
  volume       = {abs/2402.03300},
  year         = {2024},
  url          = {https://doi.org/10.48550/arXiv.2402.03300},
  doi          = {10.48550/ARXIV.2402.03300},
  eprinttype    = {arXiv},
  eprint       = {2402.03300},
  bibsource    = {dblp computer science bibliography, https://dblp.org}
}

@inproceedings{DBLP:conf/acl/AhmadianCGFKPUH24,
  author       = {Arash Ahmadian and
                  Chris Cremer and
                  Matthias Gall{\'{e}} and
                  Marzieh Fadaee and
                  Julia Kreutzer and
                  Olivier Pietquin and
                  Ahmet {\"{U}}st{\"{u}}n and
                  Sara Hooker},
  editor       = {Lun{-}Wei Ku and
                  Andre Martins and
                  Vivek Srikumar},
  title        = {Back to Basics: Revisiting REINFORCE-Style Optimization for Learning
                  from Human Feedback in LLMs},
  booktitle    = {Proceedings of the 62nd Annual Meeting of the Association for Computational
                  Linguistics (Volume 1: Long Papers), {ACL} 2024, Bangkok, Thailand,
                  August 11-16, 2024},
  pages        = {12248--12267},
  publisher    = {Association for Computational Linguistics},
  year         = {2024},
  url          = {https://doi.org/10.18653/v1/2024.acl-long.662},
  doi          = {10.18653/V1/2024.ACL-LONG.662},
  bibsource    = {dblp computer science bibliography, https://dblp.org}
}

@inproceedings{SafeRLHF,
  author       = {Josef Dai and
                  Xuehai Pan and
                  Ruiyang Sun and
                  Jiaming Ji and
                  Xinbo Xu and
                  Mickel Liu and
                  Yizhou Wang and
                  Yaodong Yang},
  title        = {Safe {RLHF:} Safe Reinforcement Learning from Human Feedback},
  booktitle    = {The Twelfth International Conference on Learning Representations,
                  {ICLR} 2024, Vienna, Austria, May 7-11, 2024},
  publisher    = {OpenReview.net},
  year         = {2024},
  url          = {https://openreview.net/forum?id=TyFrPOKYXw},
  bibsource    = {dblp computer science bibliography, https://dblp.org}
}

@inproceedings{DPO,
  author       = {Rafael Rafailov and
                  Archit Sharma and
                  Eric Mitchell and
                  Christopher D. Manning and
                  Stefano Ermon and
                  Chelsea Finn},
  editor       = {Alice Oh and
                  Tristan Naumann and
                  Amir Globerson and
                  Kate Saenko and
                  Moritz Hardt and
                  Sergey Levine},
  title        = {Direct Preference Optimization: Your Language Model is Secretly a
                  Reward Model},
  booktitle    = {Advances in Neural Information Processing Systems 36: Annual Conference
                  on Neural Information Processing Systems 2023, NeurIPS 2023, New Orleans,
                  LA, USA, December 10 - 16, 2023},
  year         = {2023},
  url          = {http://papers.nips.cc/paper\_files/paper/2023/hash/a85b405ed65c6477a4fe8302b5e06ce7-Abstract-Conference.html},
  bibsource    = {dblp computer science bibliography, https://dblp.org}
}

@article{DBLP:journals/tmlr/CasperDSGSRFKLF23,
  author       = {Stephen Casper and
                  Xander Davies and
                  Claudia Shi and
                  Thomas Krendl Gilbert and
                  J{\'{e}}r{\'{e}}my Scheurer and
                  Javier Rando and
                  Rachel Freedman and
                  Tomasz Korbak and
                  David Lindner and
                  Pedro Freire and
                  Tony Tong Wang and
                  Samuel Marks and
                  Charbel{-}Rapha{\"{e}}l S{\'{e}}gerie and
                  Micah Carroll and
                  Andi Peng and
                  Phillip J. K. Christoffersen and
                  Mehul Damani and
                  Stewart Slocum and
                  Usman Anwar and
                  Anand Siththaranjan and
                  Max Nadeau and
                  Eric J. Michaud and
                  Jacob Pfau and
                  Dmitrii Krasheninnikov and
                  Xin Chen and
                  Lauro Langosco and
                  Peter Hase and
                  Erdem Biyik and
                  Anca D. Dragan and
                  David Krueger and
                  Dorsa Sadigh and
                  Dylan Hadfield{-}Menell},
  title        = {Open Problems and Fundamental Limitations of Reinforcement Learning
                  from Human Feedback},
  journal      = {Trans. Mach. Learn. Res.},
  volume       = {2023},
  year         = {2023},
  url          = {https://openreview.net/forum?id=bx24KpJ4Eb},
  bibsource    = {dblp computer science bibliography, https://dblp.org}
}

@article{PPO,
  author       = {John Schulman and
                  Filip Wolski and
                  Prafulla Dhariwal and
                  Alec Radford and
                  Oleg Klimov},
  title        = {Proximal Policy Optimization Algorithms},
  journal      = {CoRR},
  volume       = {abs/1707.06347},
  year         = {2017},
  url          = {http://arxiv.org/abs/1707.06347},
  eprinttype    = {arXiv},
  eprint       = {1707.06347},
  bibsource    = {dblp computer science bibliography, https://dblp.org}
}

@inproceedings{GAE,
  author       = {John Schulman and
                  Philipp Moritz and
                  Sergey Levine and
                  Michael I. Jordan and
                  Pieter Abbeel},
  editor       = {Yoshua Bengio and
                  Yann LeCun},
  title        = {High-Dimensional Continuous Control Using Generalized Advantage Estimation},
  booktitle    = {4th International Conference on Learning Representations, {ICLR} 2016,
                  San Juan, Puerto Rico, May 2-4, 2016, Conference Track Proceedings},
  year         = {2016},
  url          = {http://arxiv.org/abs/1506.02438},
  bibsource    = {dblp computer science bibliography, https://dblp.org}
}

@inproceedings{Sailor,
  author       = {Foteini Strati and
                  Zhendong Zhang and
                  George Manos and
                  Ixeia S{\'{a}}nchez P{\'{e}}riz and
                  Qinghao Hu and
                  Tiancheng Chen and
                  Berk Buzcu and
                  Song Han and
                  Pamela Delgado and
                  Ana Klimovic},
  editor       = {Youjip Won and
                  Youngjin Kwon and
                  Ding Yuan and
                  Rebecca Isaacs},
  title        = {Sailor: Automating Distributed Training over Dynamic, Heterogeneous,
                  and Geo-distributed Clusters},
  booktitle    = {Proceedings of the {ACM} {SIGOPS} 31st Symposium on Operating Systems
                  Principles, {SOSP} 2025, Lotte Hotel World, Seoul, Republic of Korea,
                  October 13-16, 2025},
  pages        = {204--220},
  publisher    = {{ACM}},
  year         = {2025},
  url          = {https://doi.org/10.1145/3731569.3764839},
  doi          = {10.1145/3731569.3764839},
  bibsource    = {dblp computer science bibliography, https://dblp.org}
}

@inproceedings{Helix,
  author       = {Yixuan Mei and
                  Yonghao Zhuang and
                  Xupeng Miao and
                  Juncheng Yang and
                  Zhihao Jia and
                  Rashmi Vinayak},
  editor       = {Lieven Eeckhout and
                  Georgios Smaragdakis and
                  Kaitai Liang and
                  Adrian Sampson and
                  Martha A. Kim and
                  Christopher J. Rossbach},
  title        = {Helix: Serving Large Language Models over Heterogeneous GPUs and Network
                  via Max-Flow},
  booktitle    = {Proceedings of the 30th {ACM} International Conference on Architectural
                  Support for Programming Languages and Operating Systems, Volume 1,
                  {ASPLOS} 2025, Rotterdam, The Netherlands, 30 March 2025 - 3 April
                  2025},
  pages        = {586--602},
  publisher    = {{ACM}},
  year         = {2025},
  url          = {https://doi.org/10.1145/3669940.3707215},
  doi          = {10.1145/3669940.3707215},
  bibsource    = {dblp computer science bibliography, https://dblp.org}
}

@article{HeterMoE,
  author       = {Yongji Wu and
                  Xueshen Liu and
                  Shuowei Jin and
                  Ceyu Xu and
                  Feng Qian and
                  Z. Morley Mao and
                  Matthew Lentz and
                  Danyang Zhuo and
                  Ion Stoica},
  title        = {HeterMoE: Efficient Training of Mixture-of-Experts Models on Heterogeneous
                  GPUs},
  journal      = {CoRR},
  volume       = {abs/2504.03871},
  year         = {2025},
  url          = {https://doi.org/10.48550/arXiv.2504.03871},
  doi          = {10.48550/ARXIV.2504.03871},
  eprinttype    = {arXiv},
  eprint       = {2504.03871},
  bibsource    = {dblp computer science bibliography, https://dblp.org}
}

@inproceedings{LLMStation,
  author       = {Yongjun He and
                  Haofeng Yang and
                  Yao Lu and
                  Ana Klimovic and
                  Gustavo Alonso},
  editor       = {Deniz Altinb{\"{u}}ken and
                  Ryan Stutsman},
  title        = {Resource Multiplexing in Tuning and Serving Large Language Models},
  booktitle    = {Proceedings of the 2025 {USENIX} Annual Technical Conference, {USENIX}
                  {ATC} 2025, Boston, MA, USA, July 7-9, 2025},
  pages        = {1639--1655},
  publisher    = {{USENIX} Association},
  year         = {2025},
  url          = {https://www.usenix.org/conference/atc25/presentation/he-yongjun},
  bibsource    = {dblp computer science bibliography, https://dblp.org}
}

@article{Qwen3,
  author       = {{Qwen Team}},
  title        = {Qwen3 Technical Report},
  journal      = {CoRR},
  volume       = {abs/2505.09388},
  year         = {2025},
  url          = {https://doi.org/10.48550/arXiv.2505.09388},
  doi          = {10.48550/ARXIV.2505.09388},
  eprinttype    = {arXiv},
  eprint       = {2505.09388},
  bibsource    = {dblp computer science bibliography, https://dblp.org}
}

@article{DeepSeek-V3,
  author       = {DeepSeek{-}AI},
  title        = {DeepSeek-V3 Technical Report},
  journal      = {CoRR},
  volume       = {abs/2412.19437},
  year         = {2024},
  url          = {https://doi.org/10.48550/arXiv.2412.19437},
  doi          = {10.48550/ARXIV.2412.19437},
  eprinttype    = {arXiv},
  eprint       = {2412.19437},
  bibsource    = {dblp computer science bibliography, https://dblp.org}
}

@article{Llama3,
  author       = {{Llama Team}},
  title        = {The Llama 3 Herd of Models},
  journal      = {CoRR},
  volume       = {abs/2407.21783},
  year         = {2024},
  url          = {https://doi.org/10.48550/arXiv.2407.21783},
  doi          = {10.48550/ARXIV.2407.21783},
  eprinttype    = {arXiv},
  eprint       = {2407.21783},
  bibsource    = {dblp computer science bibliography, https://dblp.org}
}

@inproceedings{jiangdemystifying,
  title={Demystifying Cost-Efficiency in LLM Serving over Heterogeneous GPUs},
  author={Jiang, Youhe and Fu, Fangcheng and Yao, Xiaozhe and He, Guoliang and Miao, Xupeng and Klimovic, Ana and Cui, Bin and Yuan, Binhang and Yoneki, Eiko},
  booktitle={Forty-second International Conference on Machine Learning, {ICML} 2025, Vancouver, Canada, July 13-19, 2025},
  publisher    = {OpenReview.net},
  year         = {2024},
}

@article{ThunderServe,
  author       = {Youhe Jiang and
                  Fangcheng Fu and
                  Xiaozhe Yao and
                  Taiyi Wang and
                  Bin Cui and
                  Ana Klimovic and
                  Eiko Yoneki},
  title        = {ThunderServe: High-performance and Cost-efficient {LLM} Serving in
                  Cloud Environments},
  journal      = {CoRR},
  volume       = {abs/2502.09334},
  year         = {2025},
  url          = {https://doi.org/10.48550/arXiv.2502.09334},
  doi          = {10.48550/ARXIV.2502.09334},
  eprinttype    = {arXiv},
  eprint       = {2502.09334},
  bibsource    = {dblp computer science bibliography, https://dblp.org}
}

@inproceedings{HexGen,
  author       = {Youhe Jiang and
                  Ran Yan and
                  Xiaozhe Yao and
                  Yang Zhou and
                  Beidi Chen and
                  Binhang Yuan},
  title        = {HexGen: Generative Inference of Large Language Model over Heterogeneous
                  Environment},
  booktitle    = {Forty-first International Conference on Machine Learning, {ICML} 2024,
                  Vienna, Austria, July 21-27, 2024},
  series       = {Proceedings of Machine Learning Research},
  pages        = {21946--21961},
  publisher    = {{PMLR} / OpenReview.net},
  year         = {2024},
  url          = {https://proceedings.mlr.press/v235/jiang24f.html},
  bibsource    = {dblp computer science bibliography, https://dblp.org}
}

@inproceedings{DBLP:conf/euromlsys/StratiEKK24,
  author       = {Foteini Strati and
                  Paul Elvinger and
                  Tolga Kerimoglu and
                  Ana Klimovic},
  title        = {{ML} Training with Cloud {GPU} Shortages: Is Cross-Region the Answer?},
  booktitle    = {Proceedings of the 4th Workshop on Machine Learning and Systems, EuroMLSys
                  2024, Athens, Greece, 22 April 2024},
  pages        = {107--116},
  publisher    = {{ACM}},
  year         = {2024},
  url          = {https://doi.org/10.1145/3642970.3655843},
  doi          = {10.1145/3642970.3655843},
  bibsource    = {dblp computer science bibliography, https://dblp.org}
}

@inproceedings{DBLP:conf/icse/GaoHLZLLZ0WZGTY24,
  author       = {Yanjie Gao and
                  Yichen He and
                  Xinze Li and
                  Bo Zhao and
                  Haoxiang Lin and
                  Yoyo Liang and
                  Jing Zhong and
                  Hongyu Zhang and
                  Jingzhou Wang and
                  Yonghua Zeng and
                  Keli Gui and
                  Jie Tong and
                  Mao Yang},
  title        = {An Empirical Study on Low {GPU} Utilization of Deep Learning Jobs},
  booktitle    = {Proceedings of the 46th {IEEE/ACM} International Conference on Software
                  Engineering, {ICSE} 2024, Lisbon, Portugal, April 14-20, 2024},
  pages        = {96:1--96:13},
  publisher    = {{ACM}},
  year         = {2024},
  url          = {https://doi.org/10.1145/3597503.3639232},
  doi          = {10.1145/3597503.3639232},
  bibsource    = {dblp computer science bibliography, https://dblp.org}
}

@mastersthesis{zhang2024,
  title={Understanding GPU Architecture Implications on LLM Serving Workloads},
  author={Zhang, Zijian},
  year={2024},
  school={ETH Zurich}
}

@inproceedings{Metis,
  author       = {Taegeon Um and
                  Byungsoo Oh and
                  Minyoung Kang and
                  Woo{-}Yeon Lee and
                  Goeun Kim and
                  Dongseob Kim and
                  Youngtaek Kim and
                  Mohd Muzzammil and
                  Myeongjae Jeon},
  editor       = {Saurabh Bagchi and
                  Yiying Zhang},
  title        = {Metis: Fast Automatic Distributed Training on Heterogeneous GPUs},
  booktitle    = {Proceedings of the 2024 {USENIX} Annual Technical Conference, {USENIX}
                  {ATC} 2024, Santa Clara, CA, USA, July 10-12, 2024},
  pages        = {563--578},
  publisher    = {{USENIX} Association},
  year         = {2024},
  url          = {https://www.usenix.org/conference/atc24/presentation/um},
  bibsource    = {dblp computer science bibliography, https://dblp.org}
}

@inproceedings{vLLM,
  author       = {Woosuk Kwon and
                  Zhuohan Li and
                  Siyuan Zhuang and
                  Ying Sheng and
                  Lianmin Zheng and
                  Cody Hao Yu and
                  Joseph Gonzalez and
                  Hao Zhang and
                  Ion Stoica},
  editor       = {Jason Flinn and
                  Margo I. Seltzer and
                  Peter Druschel and
                  Antoine Kaufmann and
                  Jonathan Mace},
  title        = {Efficient Memory Management for Large Language Model Serving with
                  PagedAttention},
  booktitle    = {Proceedings of the 29th Symposium on Operating Systems Principles,
                  {SOSP} 2023, Koblenz, Germany, October 23-26, 2023},
  pages        = {611--626},
  publisher    = {{ACM}},
  year         = {2023},
  url          = {https://doi.org/10.1145/3600006.3613165},
  doi          = {10.1145/3600006.3613165},
  bibsource    = {dblp computer science bibliography, https://dblp.org}
}

@inproceedings{AlpaServe,
  author       = {Zhuohan Li and
                  Lianmin Zheng and
                  Yinmin Zhong and
                  Vincent Liu and
                  Ying Sheng and
                  Xin Jin and
                  Yanping Huang and
                  Zhifeng Chen and
                  Hao Zhang and
                  Joseph E. Gonzalez and
                  Ion Stoica},
  editor       = {Roxana Geambasu and
                  Ed Nightingale},
  title        = {AlpaServe: Statistical Multiplexing with Model Parallelism for Deep
                  Learning Serving},
  booktitle    = {17th {USENIX} Symposium on Operating Systems Design and Implementation,
                  {OSDI} 2023, Boston, MA, USA, July 10-12, 2023},
  pages        = {663--679},
  publisher    = {{USENIX} Association},
  year         = {2023},
  url          = {https://www.usenix.org/conference/osdi23/presentation/li-zhouhan},
  bibsource    = {dblp computer science bibliography, https://dblp.org}
}

@inproceedings{DTFM,
  author       = {Binhang Yuan and
                  Yongjun He and
                  Jared Davis and
                  Tianyi Zhang and
                  Tri Dao and
                  Beidi Chen and
                  Percy Liang and
                  Christopher R{\'{e}} and
                  Ce Zhang},
  title        = {Decentralized Training of Foundation Models in Heterogeneous Environments},
  booktitle    = {Advances in Neural Information Processing Systems 35: Annual Conference
                  on Neural Information Processing Systems 2022, NeurIPS 2022, New Orleans,
                  LA, USA, November 28 - December 9, 2022},
  year         = {2022},
  url          ={http://papers.nips.cc/paper\_files/paper/2022/hash/a37d615b61f999a5fa276adb14643476-Abstract-Conference.html},
  bibsource    = {dblp computer science bibliography, https://dblp.org}
}

@inproceedings{Alpa,
  author       = {Lianmin Zheng and
                  Zhuohan Li and
                  Hao Zhang and
                  Yonghao Zhuang and
                  Zhifeng Chen and
                  Yanping Huang and
                  Yida Wang and
                  Yuanzhong Xu and
                  Danyang Zhuo and
                  Eric P. Xing and
                  Joseph E. Gonzalez and
                  Ion Stoica},
  editor       = {Marcos K. Aguilera and
                  Hakim Weatherspoon},
  title        = {Alpa: Automating Inter- and Intra-Operator Parallelism for Distributed
                  Deep Learning},
  booktitle    = {16th {USENIX} Symposium on Operating Systems Design and Implementation,
                  {OSDI} 2022, Carlsbad, CA, USA, July 11-13, 2022},
  pages        = {559--578},
  publisher    = {{USENIX} Association},
  year         = {2022},
  url          = {https://www.usenix.org/conference/osdi22/presentation/zheng-lianmin},
  bibsource    = {dblp computer science bibliography, https://dblp.org}
}

@inproceedings{AntMan,
  author       = {Wencong Xiao and
                  Shiru Ren and
                  Yong Li and
                  Yang Zhang and
                  Pengyang Hou and
                  Zhi Li and
                  Yihui Feng and
                  Wei Lin and
                  Yangqing Jia},
  title        = {AntMan: Dynamic Scaling on {GPU} Clusters for Deep Learning},
  booktitle    = {14th {USENIX} Symposium on Operating Systems Design and Implementation,
                  {OSDI} 2020, Virtual Event, November 4-6, 2020},
  pages        = {533--548},
  publisher    = {{USENIX} Association},
  year         = {2020},
  url          = {https://www.usenix.org/conference/osdi20/presentation/xiao},
  bibsource    = {dblp computer science bibliography, https://dblp.org}
}

@inproceedings{GSLICE,
  author       = {Aditya Dhakal and
                  Sameer G. Kulkarni and
                  K. K. Ramakrishnan},
  editor       = {Rodrigo Fonseca and
                  Christina Delimitrou and
                  Beng Chin Ooi},
  title        = {{GSLICE:} controlled spatial sharing of GPUs for a scalable inference
                  platform},
  booktitle    = {SoCC '20: {ACM} Symposium on Cloud Computing, Virtual Event, USA,
                  October 19-21, 2020},
  pages        = {492--506},
  publisher    = {{ACM}},
  year         = {2020},
  url          = {https://doi.org/10.1145/3419111.3421284},
  doi          = {10.1145/3419111.3421284},
  bibsource    = {dblp computer science bibliography, https://dblp.org}
}

@inproceedings{ZeRO,
  author       = {Samyam Rajbhandari and
                  Jeff Rasley and
                  Olatunji Ruwase and
                  Yuxiong He},
  editor       = {Christine Cuicchi and
                  Irene Qualters and
                  William T. Kramer},
  title        = {ZeRO: memory optimizations toward training trillion parameter models},
  booktitle    = {Proceedings of the International Conference for High Performance Computing,
                  Networking, Storage and Analysis, {SC} 2020, Virtual Event / Atlanta,
                  Georgia, USA, November 9-19, 2020},
  pages        = {20},
  publisher    = {{IEEE/ACM}},
  year         = {2020},
  url          = {https://doi.org/10.1109/SC41405.2020.00024},
  doi          = {10.1109/SC41405.2020.00024},
  bibsource    = {dblp computer science bibliography, https://dblp.org}
}

@article{Megatron-LM,
  author       = {Mohammad Shoeybi and
                  Mostofa Patwary and
                  Raul Puri and
                  Patrick LeGresley and
                  Jared Casper and
                  Bryan Catanzaro},
  title        = {Megatron-LM: Training Multi-Billion Parameter Language Models Using
                  Model Parallelism},
  journal      = {CoRR},
  volume       = {abs/1909.08053},
  year         = {2019},
  url          = {http://arxiv.org/abs/1909.08053},
  eprinttype    = {arXiv},
  eprint       = {1909.08053},
  bibsource    = {dblp computer science bibliography, https://dblp.org}
}

@inproceedings{GPipe,
  author       = {Yanping Huang and
                  Youlong Cheng and
                  Ankur Bapna and
                  Orhan Firat and
                  Dehao Chen and
                  Mia Xu Chen and
                  HyoukJoong Lee and
                  Jiquan Ngiam and
                  Quoc V. Le and
                  Yonghui Wu and
                  Zhifeng Chen},
  editor       = {Hanna M. Wallach and
                  Hugo Larochelle and
                  Alina Beygelzimer and
                  Florence d'Alch{\'{e}}{-}Buc and
                  Emily B. Fox and
                  Roman Garnett},
  title        = {GPipe: Efficient Training of Giant Neural Networks using Pipeline
                  Parallelism},
  booktitle    = {Advances in Neural Information Processing Systems 32: Annual Conference
                  on Neural Information Processing Systems 2019, NeurIPS 2019, December
                  8-14, 2019, Vancouver, BC, Canada},
  pages        = {103--112},
  year         = {2019},
  url          = {https://proceedings.neurips.cc/paper/2019/hash/093f65e080a295f8076b1c5722a46aa2-Abstract.html},
  bibsource    = {dblp computer science bibliography, https://dblp.org}
}

@inproceedings{Gandiva,
  author       = {Wencong Xiao and
                  Romil Bhardwaj and
                  Ramachandran Ramjee and
                  Muthian Sivathanu and
                  Nipun Kwatra and
                  Zhenhua Han and
                  Pratyush Patel and
                  Xuan Peng and
                  Hanyu Zhao and
                  Quanlu Zhang and
                  Fan Yang and
                  Lidong Zhou},
  editor       = {Andrea C. Arpaci{-}Dusseau and
                  Geoff Voelker},
  title        = {Gandiva: Introspective Cluster Scheduling for Deep Learning},
  booktitle    = {13th {USENIX} Symposium on Operating Systems Design and Implementation,
                  {OSDI} 2018, Carlsbad, CA, USA, October 8-10, 2018},
  pages        = {595--610},
  publisher    = {{USENIX} Association},
  year         = {2018},
  url          = {https://www.usenix.org/conference/osdi18/presentation/xiao},
  bibsource    = {dblp computer science bibliography, https://dblp.org}
}

@inproceedings{Hyperband,
  author       = {Lisha Li and
                  Kevin G. Jamieson and
                  Giulia DeSalvo and
                  Afshin Rostamizadeh and
                  Ameet Talwalkar},
  title        = {Hyperband: Bandit-Based Configuration Evaluation for Hyperparameter
                  Optimization},
  booktitle    = {5th International Conference on Learning Representations, {ICLR} 2017,
                  Toulon, France, April 24-26, 2017, Conference Track Proceedings},
  publisher    = {OpenReview.net},
  year         = {2017},
  url          = {https://openreview.net/forum?id=ry18Ww5ee},
  bibsource    = {dblp computer science bibliography, https://dblp.org}
}

@inproceedings{SHA,
  author       = {Kevin G. Jamieson and
                  Ameet Talwalkar},
  editor       = {Arthur Gretton and
                  Christian C. Robert},
  title        = {Non-stochastic Best Arm Identification and Hyperparameter Optimization},
  booktitle    = {Proceedings of the 19th International Conference on Artificial Intelligence
                  and Statistics, {AISTATS} 2016, Cadiz, Spain, May 9-11, 2016},
  series       = {{JMLR} Workshop and Conference Proceedings},
  volume       = {51},
  pages        = {240--248},
  publisher    = {JMLR.org},
  year         = {2016},
  url          = {http://proceedings.mlr.press/v51/jamieson16.html},
  bibsource    = {dblp computer science bibliography, https://dblp.org}
}

@article{DEAP, 
    author    = " F\'elix-Antoine Fortin and Fran\c{c}ois-Michel {De Rainville} and Marc-Andr\'e Gardner and Marc Parizeau and Christian Gagn\'e ",
    title     = { {DEAP}: Evolutionary Algorithms Made Easy },
    pages    = { 2171--2175 },
    volume    = { 13 },
    month     = { jul },
    year      = { 2012 },
    journal   = { Journal of Machine Learning Research }
}

@article{DBLP:journals/jgo/SoperWC04,
  author       = {Alan J. Soper and
                  Chris Walshaw and
                  Mark Cross},
  title        = {A Combined Evolutionary Search and Multilevel Optimisation Approach
                  to Graph-Partitioning},
  journal      = {J. Glob. Optim.},
  volume       = {29},
  number       = {2},
  pages        = {225--241},
  year         = {2004},
  url          = {https://doi.org/10.1023/B:JOGO.0000042115.44455.f3},
  doi          = {10.1023/B:JOGO.0000042115.44455.F3},
  bibsource    = {dblp computer science bibliography, https://dblp.org}
}

@article{DBLP:journals/tc/BuiM96,
  author       = {Thang Nguyen Bui and
                  Byung Ro Moon},
  title        = {Genetic Algorithm and Graph Partitioning},
  journal      = {{IEEE} Trans. Computers},
  volume       = {45},
  number       = {7},
  pages        = {841--855},
  year         = {1996},
  url          = {https://doi.org/10.1109/12.508322},
  doi          = {10.1109/12.508322},
  bibsource    = {dblp computer science bibliography, https://dblp.org}
}

@article{DBLP:journals/compsys/HintonN87,
  author       = {Geoffrey E. Hinton and
                  Steven J. Nowlan},
  title        = {How Learning Can Guide Evolution},
  journal      = {Complex Syst.},
  volume       = {1},
  number       = {3},
  year         = {1987},
  url          = {http://www.complex-systems.com/abstracts/v01\_i03\_a06.html},
  bibsource    = {dblp computer science bibliography, https://dblp.org}
}

@article{rota1964number,
  title={The number of partitions of a set},
  author={Rota, Gian-Carlo},
  journal={The American Mathematical Monthly},
  volume={71},
  number={5},
  pages={498--504},
  year={1964},
  publisher={Taylor \& Francis}
}

@article{baldwin1896new,
  title={A new factor in evolution},
  author={Baldwin, J Mark},
  journal={Adaptive individuals in evolving populations: Models and algorithms},
  pages={59--80},
  year={1896},
  publisher={Addison-Wesley Reading, MA}
}
